# Engineering reaction-diffusion networks with properties of neural tissue

**Thomas Litschel,[a] Michael M. Norton,[a] Vardges Tserunyan,[a] Seth Fraden*[a]**

**[a]Department of Physics, Brandeis University, Waltham, MA 02453.**
***Corresponding Author. Email: fraden@brandeis.edu**

**We present an experimental system of networks of coupled non-linear chemical reactors, which we theoretically model within a reaction-diffusion framework. The networks consist of patterned arrays of diffusively coupled nanoliter-scale reactors containing the Belousov-Zhabotinsky (BZ) reaction. Microfluidic fabrication techniques are developed that provide the ability to vary the network topology, the reactor coupling strength and offer the freedom to choose whether an arbitrary reactor is inhibitory or excitatory coupled to its neighbor. This versatile experimental and theoretical framework can be used to create a wide variety of chemical networks. Here we design, construct and characterize chemical networks that achieve the complexity of central pattern generators (CPGs), which are found in the autonomic nervous system of a variety of organisms.**

## INTRODUCTION

Central pattern generators are a class of neural networks found in the autonomous nervous system that perform cyclic functions and govern motions such as swimming, walking, and the peristaltic motion of the digestive system. The networks produce rhythmic pulses distributed in space that coordinate muscle contractions (Fig. 1). Minimal models treat individual neurons as self-driven non-linear oscillators that when coupled together produce complex spatio-temporal patterns.[1, 2] Turing recognized that chemical reactions were capable of producing self-driven oscillators and elucidated the conditions in which diffusively coupled chemical networks exhibit spontaneous spatio-temporal pattern formation.[3] Guided by this insight, we seek to leverage the self-organizing properties of reaction-diffusion systems to engineer a synthetic substrate that emulates the autonomous spatio-temporal patterns exhibited by central-pattern generators. In this paper, we describe the design and fabrication of a network of coupled chemical oscillators that generates the spatio-temporal pattern produced by a swimming lamprey (Fig. 1).

As the name implies, reaction-diffusion networks have two components. The term *reaction* refers to the chemistry. As a first step of the design process we, like Turing, idealize the chemical kinetics as taking place inside a single reactor, disregarding spatial extent. The term *diffusion networks* refers to the physics with a focus on the spatial structure and inter-connectivity of the network as characterized by (*i*) the topology of the network, the (*ii*) boundary and (*iii*) initial conditions, (*iv*) the volume of each reactor, (*v*) the coupling strength, and (*vi*) whether the coupling is of an inhibitory or excitatory nature. To enable the rational engineering of reaction-diffusion networks it is important to have control over each of these factors.

We utilize the Belousov-Zhabotinsky (BZ) reaction, a non-linear chemical oscillator. The BZ reaction is a metal ion-catalyzed oxidation of an organic substrate, in which chemical concentrations of intermediate reactants and products oscillate over time.[4, 5] The simplest description of this oscillatory reaction is with two principle components; an autocatalytic activator that drives the catalyst to the oxidized state and an inhibitor that returns the catalyst to the reduced state.[6-8] The ability of the BZ chemistry to operate in either an oscillatory/limit-cycle or excitable regime makes it a prime analogue for neurons. The parallel between the two systems has been appreciated and leveraged to produce neuro-mimetic systems by linking together stirred reactors via reagent pumps.[9] Pattern formation in larger networks of photo-chemically linked catalyst-laden beads and gels have been observed as well.[10-12] While these examples demonstrate the strength of BZ as a model oscillator, the systems do not possess inherent dynamics that are independent of the electrical, mechanical and optical controllers responsible for inter-oscillator coupling and are therefore not suitable for the creation of autonomous materials with intrinsic pattern formation.

In recent years, the Belousov-Zhabotinsky (BZ) reaction encapsulated in emulsions of droplets with diameters in the 100 μm range has emerged as a powerful experimental system to study reaction-diffusion networks. The state of the art of BZ networks up to the year 2017 has been reviewed in an article in *Lab on a Chip*; "Chemical communication and dynamics of droplet emulsions in networks of Belousov–Zhabotinsky micro-oscillators produced by microfluidics".[13] In two-dimensions (2-D) it is possible to create networks of varied topology and, using a photo-sensitive catalyst, set initial and boundary conditions.[14] Thus, diffusion network properties (*i* - *iii*) have been achieved. However, while small networks of drops of different volume have been formed in multiple emulsions, network topology and volume cannot be independently varied using this method.[15] Additionally, control of coupling strength is limited and means are lacking for specifying whether a particular link between drops in BZ emulsions is excitatory or inhibitory. Thus, properties (*iv* - *vi*) remain to be developed. The aim of this work is to create BZ reaction-diffusion networks in which all the diffusion network properties, (*i* - *vi*), can be realized and to exploit these properties to engineer non-trivial reaction-diffusion networks.

## EXPERIMENTAL METHODS

Surfactant stabilized emulsions consisting of aqueous BZ droplets containing the oscillatory BZ reaction in a continuous phase of oil allow only a subset of the BZ reactants to diffuse from drop to drop in these water-in-oil emulsions. Depending on the BZ conditions and how the BZ chemicals react with the oil and surfactant, coupling is either dominated by bromine, $Br_2$, an apolar molecule, which in the simplified classification, acts as an inhibitor,[14, 16-22] or coupling is dominated by excitatory species.[19, 23] Other ways to induce excitatory coupling are to immobilize the catalyst in specific sites and then allow all the BZ reagents minus the catalyst to diffuse over the sites.[24, 25] Thus, while there are there are several systems which allow either all the links in a reaction-diffusion network to be excitatory or all the links to be inhibitory, there are no BZ networks for which each individual link can be specified as being either excitatory or inhibitory.[13]

To create a network in which each link can either be excitatory or inhibitory, we replace the liquid oil used previously in the BZ emulsion system[14, 16-21] with the solid elastomer polydimethylsiloxane (PDMS). The oils used in emulsions and the elastomer PDMS have similar dielectric constants. Additionally, only 10% of the PDMS is cross-linked. Therefore, at a molecular scale, the majority of the elastomer resembles the fluid oil and consequently, the permeation and diffusion constants of BZ chemicals in PDMS and oil are similar.[18, 26, 27] Similarly to BZ emulsions, we found that adjacent BZ wells that were separated from each other by a wall of PDMS (Fig. 2A) were inhibitory coupled. Excitatory coupling was accomplished by building thin channels between wells (Fig. 2B). Wells connected by narrow channels allow the propagation of chemical waves, a phenomena characteristic of bulk, excitable BZ solutions.[26] Substantive advantages of PDMS confinement of BZ reactions over that of previous emulsion methods are that extensions of soft lithography methods greatly eases compartmentalization of BZ solution in PDMS wells of arbitrary size, shape and location in 2-D and allows selection of coupling type.

To pattern these nanoliter-scale wells into thin sheets of PDMS, we optimized existing microfluidic fabrication techniques[14, 28] (Fig S1A). Crucial to the fabrication of the networks is to minimize the layer of PDMS at the bottom of each well (Fig. 2). The bromine produced by the BZ reaction partitions into this space,[26, 27] and the reaction stops if the ratio of PDMS to BZ exceeds about 5:1,[14] because a critical concentration of bromine is necessary to sustain the reaction. Therefore, we fabricate PDMS sheets in which the thickness of the bottom of the well is less than 1 micron (Fig. S2). After immersing the wells in BZ solution, we seal the filled BZ compartments air-tight with a glass lid (Fig. S3) to prevent nucleation of $CO_2$ bubbles, which are a nuisance that often plague BZ experiments.[29] Manufacturing protocols for high-yield device fabrication are documented in movies S6 and S7 and in the ESI.

Our engineering design is guided by considering reaction-diffusion systems as being described by partial differential equations whose spatio-temporal behavior is a function of boundary and initial conditions. To control (ii) boundary and (iii) initial conditions, we include the photosensitive catalyst tris-(bipyridine)-ruthenium(II) ($[Ru(bpy)_3]^{2+}$) in the BZ solution. Essentially, when illuminated with blue light, a $[Ru(bpy)_3]^{2+}$ - BZ solution is prevented from oscillating and is set in partially oxidized steady state.[30] We utilize a homemade Programmable Illumination Microscope (PIM)[31] for a light source. This microscope has two illumination arms; one to control the chemical reaction and one to observe the oxidation state. The control arm consists of a computer projector capable of projecting blue light with wavelength centered at 450 nm at arbitrary locations over the entire field of view and allows the intensity of each pixel to be varied in time.[14, 31] The observation arm utilizes Kohler illumination at 510 nm to maximize the contrast of the ferroin dye that indicates the oxidation state of the $[Ru(bpy)_3]^{2+}$ catalyst.[14, 18]

To establish defined boundary conditions, we fabricate an additional structure (a moat) in the PDMS containing BZ that surrounds, but is not directly connected to the network of interest (Fig. 3A). The BZ chemical composition is initially the same in all structures, but using the PIM to illuminate the BZ in the moat with sufficient light moves the reaction off the limit cycle, fixing the reaction in a stationary state of constant chemical concentration, including a fixed bromine concentration. This establishes a well-controlled, time-invariant boundary condition consisting of constant chemical concentrations in a region surrounding the network under study.[14, 18, 21] The non-illuminated wells continue to oscillate, while the surrounding illuminated moat isolates the network from the environment. Without the light controlled boundary conditions, a small array of BZ wells, such as illustrated in Fig. 3 would exhibit transient behavior due to mass transfer of bromine from the well into the surrounding material beyond the moat, which constitutes a much larger volume than the 1 nl wells containing the BZ solution.

We impose initial conditions by illuminating each well individually and then extinguishing the light on different wells. If the light is removed from all the wells at the same time, then the wells begin oscillating in-phase. If the light is removed at different times, then the oscillations in different wells have a phase shift proportional to the time of light removal. Once the light is removed, the BZ reaction commences and rapidly reaches the limit cycle in the reduced state. Oscillations begin approximately half an oscillation period after removal of light.

## RESULTS

### Spatial control of wave speed in linear arrays of wells

We constructed linear arrays of wells connected by small channels in the manner shown in Fig. 2B. Fig. 4 shows such a design. Well dimensions were held constant 100 μm x 100 μm x 110 μm (LxWxH) while the widths (W) and heights (H) of the connecting channels were varied. We explore the engineering of excitatory connections by investigating the dependence of wave speed on geometry. We expected that the abrupt change in channel width would induce a curvature in the wave front as it emanates from the small channel into the large well, which causes the wave to slow down according to previous experiments that were explained theoretically by the Eikonal equation.[26, 32-34] This well established phenomena occurs when an isolated reaction-diffusion wave front emerges from a small channel into a large volume and thereby acquires a curvature (circular in 2-D and spherical in 3-D); curved reaction-diffusion fronts dilute the activator responsible for propagating the reaction front, subsequently reducing wave speed. As the wave front advances into the well and interacts with the wall, the curvature reduces and concomitantly the wave speed increases. Instead, as described below, we found the opposite behavior occurred; wave velocity increased upon exiting a small channel into a larger one.

To set the direction of wave propagation, we illuminate all but one of the wells on one end of the 1-D array. The un-illuminated well oscillates spontaneously and serves as a “pacemaker” (see Fig. 3B and 4B). The light intensity is strong enough to suppress spontaneous oscillations in the illuminated wells, but weak enough such that they are still excitable and oxidize when they receive activator from a neighboring well. Such a technique has been used to study wave-propagation in continuous media.[35] Using a pacemaker allows for repeatable experiments, but is also a necessary control feature for the biological central pattern generator neural network shown in the introduction. We program pacemakers in adjacent channels to be located at opposite ends of the channels (Fig. 4C), which causes the waves to propagate in opposite directions. Comparing the resulting wave speeds controls against possible undesirable experimental gradients, such as uneven illumination. With each experiment, we include a continuous channel control, (Fig. 4C, right).

A space-time plot, Fig. 5B, shows the discretized nature of the wave as it moves from well to well. Fig. 5B conveys that the wave speed inside a well is much greater than the speed in the narrow channel between wells. Movie S1 shows a recording with 3 simultaneous, but independent experiments; the wave speeds of each propagating wave are shown in Fig. S4A. The system-average wave speeds of each experiment are constant over time, establishing that the experiments have reached steady state. As an aside, comparison with Fig. S4B reveals a similar dependency on geometry for experiments without a pacemaker well created by light. However, these experiments lacked spatial control over the wave origin and we observed waves would simultaneously randomly initiate in multiple wells.

The main experimental trend is that wave speed increases with increasing cross section of the connecting channels (Fig. 6A). The Eikonal equation predicts that the speed of isolated plane waves should be the same in long uniform channels independent of width, as is borne out in our finite element simulations (Fig. S5). However, Fig. 7 shows that the wave speed is fast in the wide uniform-width well and slow in the narrow uniform-width connection. This is because the periodicity of the spacing of the wells is close to the width of the wave front (movie S2). Thus the assumption of isolated plane waves breaks down and the Eikonal equation prediction no longer holds.[33, 34] In contrast, Ginn et. al.[26] observed that the wave speed slowed down when transitioning from a narrow to wide channel, consistent with the Ekional equation. This is because the length of the channels was much greater than the width of the wave front, closer to the idealized case described by the Eikonal equation.

We compare our results to 2-D finite element method (FEM) simulations to better understand the relationship between geometry and wave speed. The green curve in Fig. 6A shows the calculated wave speed as a function of channel width; the Field-Korös-Noyes (FKN)[5] model over predicts the wave speed by a factor of 2.5, but our model captures the overall geometric dependence. Movie S2 shows FEM simulations of wave propagation through a linear array for the case of a 10 μm connecting channel (color map corresponds to the oxidized catalyst concentration).

For very small channel sizes, we found that only every other wave propagates to the end of the array. Furthermore, the distance the disrupted waves propagate before termination shows a pattern that repeats every 4 cycles (Fig. 6 B and C, and movie S3). These periodic behaviors are sensitive to the amount of light we apply to suppress free oscillations. Similar periodic phenomena have been found experimentally and theoretically in continuous BZ systems.[36, 37] Additional quantitative information about the period of oscillations as a function of time in experiment and the spatial dependency of propagation speed in experiment and theory can be found in Fig. S4 (C to F).

Given the importance of boundary conditions for reaction-diffusion systems, we calculated FEM-predicted wave speed for bromine-permeable walls in order to model the experimentally utilized PDMS channels. For comparison, we also employed no-flux walls, which model glass or silicon devices, to explore the role of boundary conditions on wave speed. We were surprised to find only a small difference in wave speed of at most a few percent between the bromine permeable and bromine impermeable channels (Fig. S4G).

Previous researchers noticed how bromine-permeable PDMS affects BZ chemical dynamics. Buskohl and Vaia[27] used PDMS to control initial conditions. By placing PDMS walls in contact with a BZ gel, they generated a bromine sink that disinhibited a specific region. This creates a pacemaker by locally absorbing bromine. However, the bromine concentration in the PDMS will increase with time until the bromine reaches the outside of the PDMS and encounters a new boundary condition. Thus, PDMS on its own does not constitute a time-independent boundary condition. This is in contrast to the photo inhibited moat (Fig. 3A); as long as the light shone on the moat is constant, the chemical composition of the moat will also be constant.

Ginn et. al.[26] measured the wave speed in a narrowing PDMS channel and observed that the speed increased as the channel width decreased. This phenomenon was explained as arising from the permeation of the inhibitor bromine into the PDMS, thereby disinhibiting the BZ reaction. Narrow channels have more disinhibition that wide channels due to the larger surface to volume ratio. In contrast, in our wave speed experiments and simulations, the wave speed was slower in the narrow section (Fig. 5 and 6). Additionally, in our simulations the wave speed was the same for bromine permeable and impermeable channels (Fig S4G). The differences between previous observations of wave speed in PDMS channels[26] and our work is that we employ spatially modulated channels whose wave length is approximately equal to the width of the wave front. In such cases, it has been shown that the wave speed is greatly reduced, even to the extent of blocking wave propagation.[33, 34] This suggests that the wave speed in the case we studied is controlled more by channel geometry than by channel composition. This has important engineering consequences for designing autonomous networks in which wave speeds vary in different locations because it is much easier to lithographically fabricate networks with spatially variable geometry than to engineer networks with spatially variable chemical composition.

## Central pattern generator

Thus far in this paper we've established the ability to control initial and boundary conditions, wave speed, and create pacemaker nodes. We conclude by adding an additional feature: well-coupling through the inhibitor (Fig. 2A). In previous works, we explored inhibitor-coupling at length using fluorinated oils as the interstitial medium; here, we implement inhibitor-coupling for the first time using PDMS elastomer and demonstrate that both materials function identically in that the dynamic attractor for two oscillators is $180^{\circ}$ out-of-phase, also known as anti-phase synchrony. We use these engineering principles in concert to design and fabricate a central pattern generator circuit inspired by the lamprey and illustrated in Fig. 1.

This neural network is responsible for autonomous muscle control in a variety of organisms.[38, 39] The CPG's function is to initiate waves of muscle contractions that alternatively travel down the left and right sides of the spinal column, giving rise to the sinuous motion of the lamprey as illustrated in Fig. 1A.[2, 38] The minimal model that produces this behavior is that of two parallel chains of oscillators, shown in Fig. 1D.[1] Fig. 8A shows the BZ implementation in which structure implies function. It is comprised of two identical linear arrays of BZ oscillators which play the role of the neurons along the lamprey's spine. Wells within each linear array are activator-coupled through direct connections (Fig. 2B) and are designed to propagate a chemical wave in one direction as described in the previous section (Fig. 4). In order that the left and right arrays synchronize with a 180° phase difference, the two linear arrays are fabricated side-by-side to create the inhibitory interaction through the PDMS (Fig. 2A). As before, illumination is used to control boundary and initial conditions, as well as to create a pacemaker for each column (Fig. 8A).

To demonstrate the robustness of the design of the CPG shown in Fig. 8A, we purposely initiate the system as close to in-phase synchrony as experimentally possible, which is an initial condition as far away from the designed anti-phase dynamical attractor as possible. The space-time plot in Fig. 8B is recorded after the network is in steady state and corresponds closely to the target spatio-temporal pattern of the central pattern generator shown Fig. 1C. The phase difference between the left (red) and right (blue) halves of the central pattern generator vs. time shown in Fig. 8C demonstrates that the phase difference between linear arrays grows in time until the twin sides of the CPG phase lock near 180°, which is the designed dynamical attractor; additional experimental observations are shown in the supplement, Fig. S6. Movie S5 shows the FEM simulations that correspond to the experiments shown in Figs. 8C and movie S4. Fig. 8D shows the steady state phase difference between the linear arrays, for 33 separate experiments. Variations away from the desired attractor are due to heterogeneities in the system, our models reproduce this behavior when the boundary conditions are not equal on each side of the arrays. Notably, there is an additional small peak at $0^{o}$ phase shift, indicative of a second basin of attraction in the network. Our numerical models predict that the in-phase attractor is both small and shallow. The occurrence of this state is therefore controlled by both the initial condition and boundary conditions of the system. In spite of initiating the CPG with the least favorable initial condition possible, the CPG attained the designed attractor 85% of the time, demonstrating design robustness. While the great majority of experiments achieve near anti-phase synchrony, we observe large variability in the time of approach to steady state. We discuss these observations and compare to model results in the supplement, Fig. S6 – Fig. S8.

## Conclusions

Neuronal networks make possible the rich set of behaviors exhibited by animal life. At the coarsest level of description, neurons are non-linear oscillators that when coupled together in tissue through excitatory and inhibitory connections give rise to complex spatio-temporal patterns. When organized, these patterns are capable of processing and storing sensory information, and actuating musculature. Extrapolating from this general definition of a neuronal network, we posit these dynamics can be captured on an abiologic reaction-diffusion platform.

Here, we demonstrated the ability to engineer a synthetic reaction-diffusion network designed to generate spatio-temporal patterns equivalent to that of a biologically neural network, the central pattern generator. We employed the well known oscillatory Belousov-Zhabotinsky reaction and developed methods to create diffusively coupled networks over which we designed (*i*) the topology of the network, the (*ii*) boundary and (*iii*) initial conditions, (*iv*) the volume of each reactor, (*v*) the coupling strength, and (*vi*) whether the coupling is of an inhibitory or excitatory nature. It is important to note that the engineering principles we identified are general and can be applied to other oscillatory reaction-diffusion systems.

A potential application of reaction-diffusion based central pattern generators is to the field of soft robotics. One goal of soft robotics is to create complex behaviors that are inherent to the material itself and therefore require little to no computational overhead to perform; this is in contrast to classical forms of autonomous electro-mechanical systems for which control schemes must not only be carefully preconceived, but also executed at continuous expense. While the embedded control paradigm may offer less flexibility, the analog nature may contribute to robustness when the behaviors are phenotypes of the system's stable attractors. For example, the lamprey may not be capable of arbitrary swimming gaits, but the sinusoidal gait is inherent to the nervous system and musculature, which contributes to its robustness. Similarly, the dual row BZ-CPG (Fig. 8) readily "finds" a complex stable attractor from uniform initial conditions. In creating this device, we confront a new design paradigm in which information propagation and control are fully intertwined. Our future design goals include creating networks with multiple attractors that can be toggled by perturbing a small subset of the network.[40] The subsequent challenge will be transferring the information coded in the dynamical spatiotemporal patterns of compartmentalized chemical reactors into a targeted mechanical response, such as by coupling the central pattern generator demonstrated here to chemomechanical gels that contract in concert with the BZ reaction.[41, 42]

**Acknowledgements:**

We thank Xiaotong Geng, Rémi Boros, Matthew Cambria and Ethan Chan for their help as undergraduate researchers during the initial phase of the project. **Funding:** We acknowledge financial support from NSF DMREF-1534890, U. S. Army Research Office 68317-EG, and the microfluidics facility of the NSF MRSEC DMR-1420382. **Author Contributions:** S.F. and T.L. developed the experimental setup and designed the experiments. T.L. and V.T. fabricated microfluidic devices. T.L. and V.T. performed experiments. T.L. and M.N. analyzed the data. M.N. developed the theory and performed computer simulations. T.L., S.F. and M.N. wrote the manuscript. T.L. and M.N. developed the figures. **Conflicts of Interest:** There are no conflicts to declare.

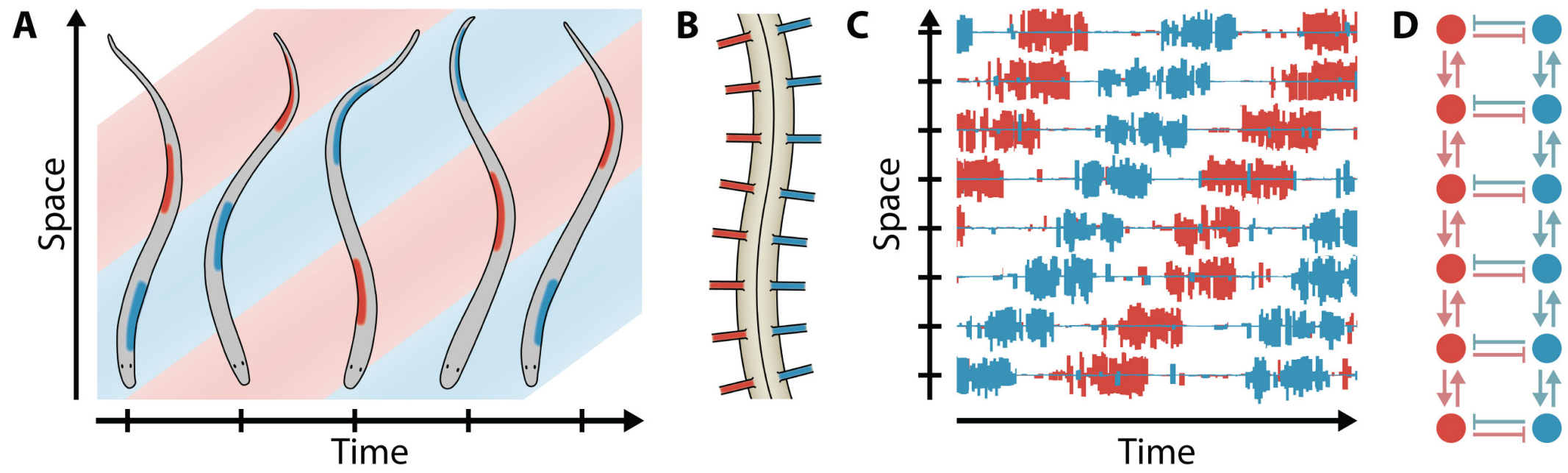

**Fig. 1. Example of a biological central pattern generator.**[1] (**A**) Lampreys swim by generating waves of transverse displacement down their spinal column. Red areas indicate regions of muscle contraction left of the spinal cord, blue indicates regions of muscle contraction right of the spinal cord. (**B**) Schematic diagram of an exposed lamprey spinal cord with ventral roots. Each vertebra has a pair of neurons, situated to the left and right of the column. (**C**) Schematic time trace of electrical activity for each of the corresponding pairs of ventral roots shown in (B). In red (blue) traces show the voltage of neurons on the left (right) side of the spinal cord. Neurons on opposite sides of a vertebra fire with a relative phase shift of 180°, while neurons on the same side fire with a small phase lag between neighboring vertebrae, producing the displacement wave shown in (A). (**D**) Schematic of lamprey central pattern generator realized as a reaction-diffusion network. Each circle represents an oscillating chemical reaction. Links with nails (arrows) indicate inhibitory (excitatory) coupling.

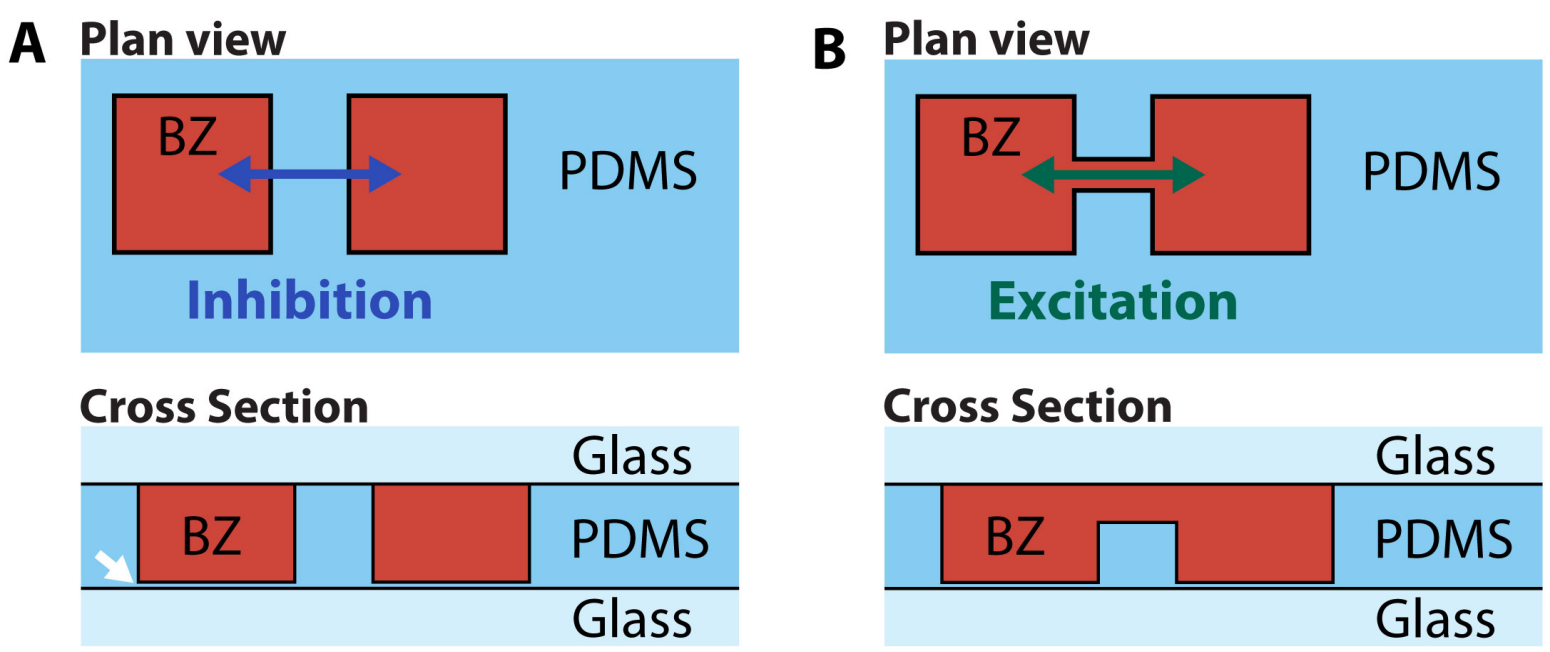

**Fig. 2. Inhibitory and excitatory coupled BZ reactors.** Plan and section views of two compartments that are designed to be inhibitory coupled (**A**) or excitatory coupled (**B**). BZ solution in red, PDMS in blue, glass in light blue. Diffusional coupling is indicated by double-arrows. It is important to minimize the thickness of PDMS below each well (white arrow) during fabrication (see Fig. S2).

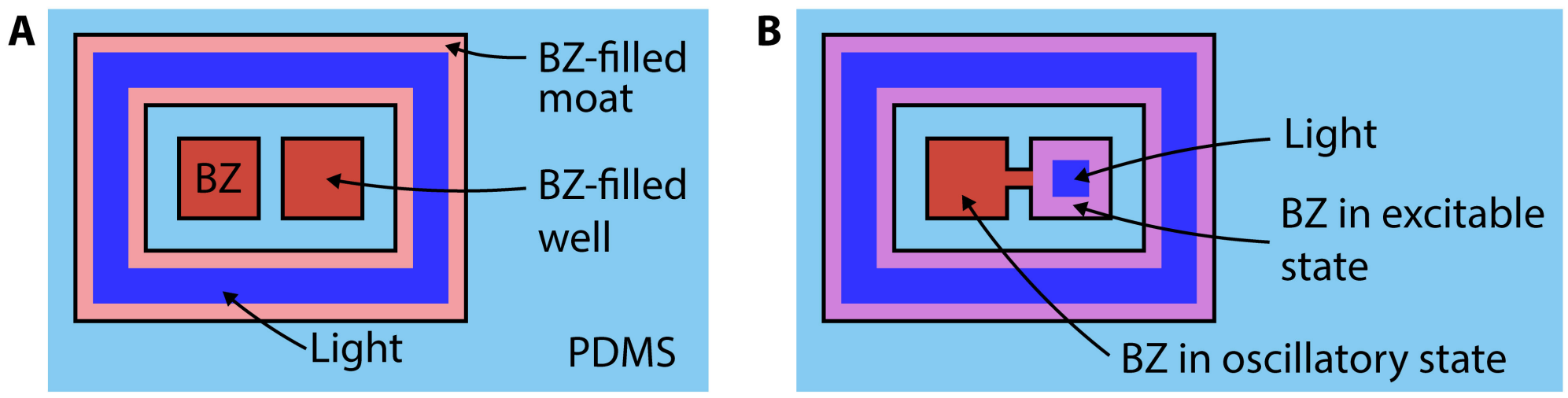

**Fig. 3. Boundary conditions and pacemakers.** (**A**) Two BZ filled wells (red) surrounded by a BZ filled moat. Blue light (blue frame) sets the BZ in the moat in an oxidized stationary state (light red). (**B**) Two BZ filled wells with a channel connection, surrounded by a BZ filled moat. The right well and the moat are exposed to light (blue square and blue frame, respectively). If the light intensity is low enough, the BZ reaction is set into a non-oscillatory, but excitable state (purple). Oscillations in the surrounding moat are suppressed; but activator from the oscillating left well ("pacemaker well"), can diffuse through the channel and oxidize the BZ in the right well.

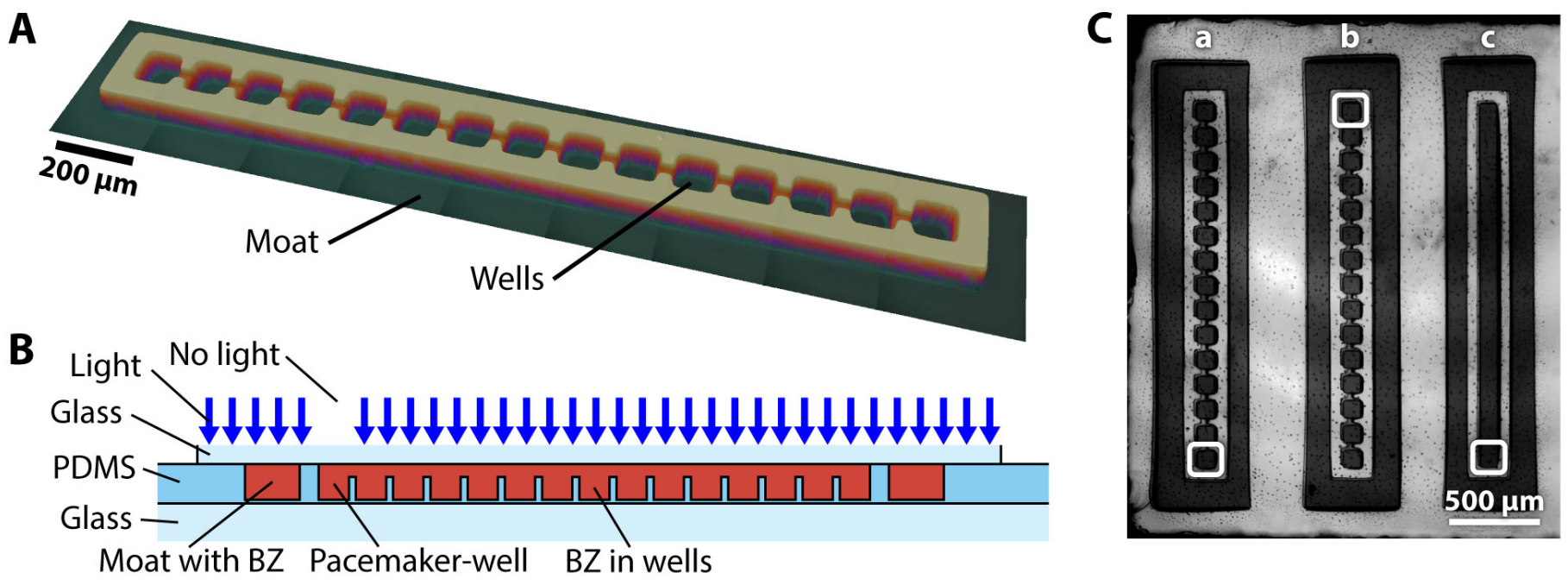

**Fig. 4. 1-D arrays of wells.** (**A**) Image of a row of wells fabricated in PDMS interconnected by channels taken with an optical profilometer and rendered in false color. (**B**) Schematic of cross section along the chain of wells. Dimensions not to scale. (**C**) Photograph of PMDS wells filled with BZ solution and surrounded by moats to establish constant chemical boundary conditions. The left (a) and the middle array (b) are geometrically identical arrays consisting of fifteen wells connected by channels. On the right (c) is a continuous chamber. Pacemaker wells, uninhibited by light, are located on alternate ends of each array and are indicated by white boxes. Movie S3 shows an experiment with propagating waves in each array.

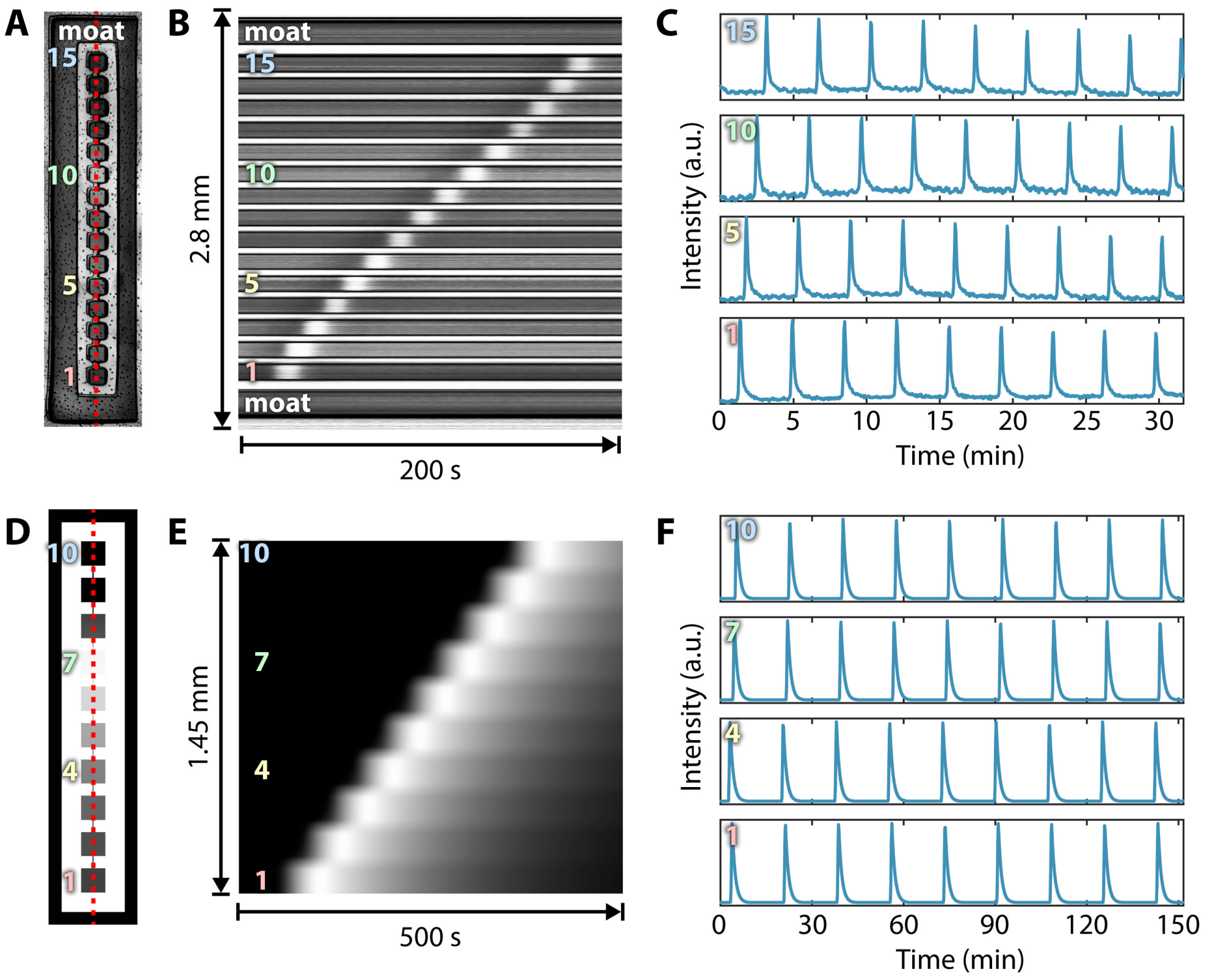

**(C)**

**Fig. 5. Waves in 1-D arrays of wells.** (**A**) Photograph of a linear array of 15 connected wells filled with BZ solution. Well *1* is the un-illuminated pacemaker well (see Fig. 4B). Shown is the left array in movie S1. (**B**) Space-time plot along the red dotted line in (A), showing a single propagating wave. The traveling wave pauses upon exiting each well. The intensity is proportional to the concentration of oxidized catalyst. (**C**) Intensity traces of 4 selected wells. Numbers in the top left corners refer to position of the wells in the array. (**D** to **F**) 2-D finite element simulation of the experiment shown in (A to C). Shown is the concentration of oxidized catalyst (white: high concentration; black: low concentration).

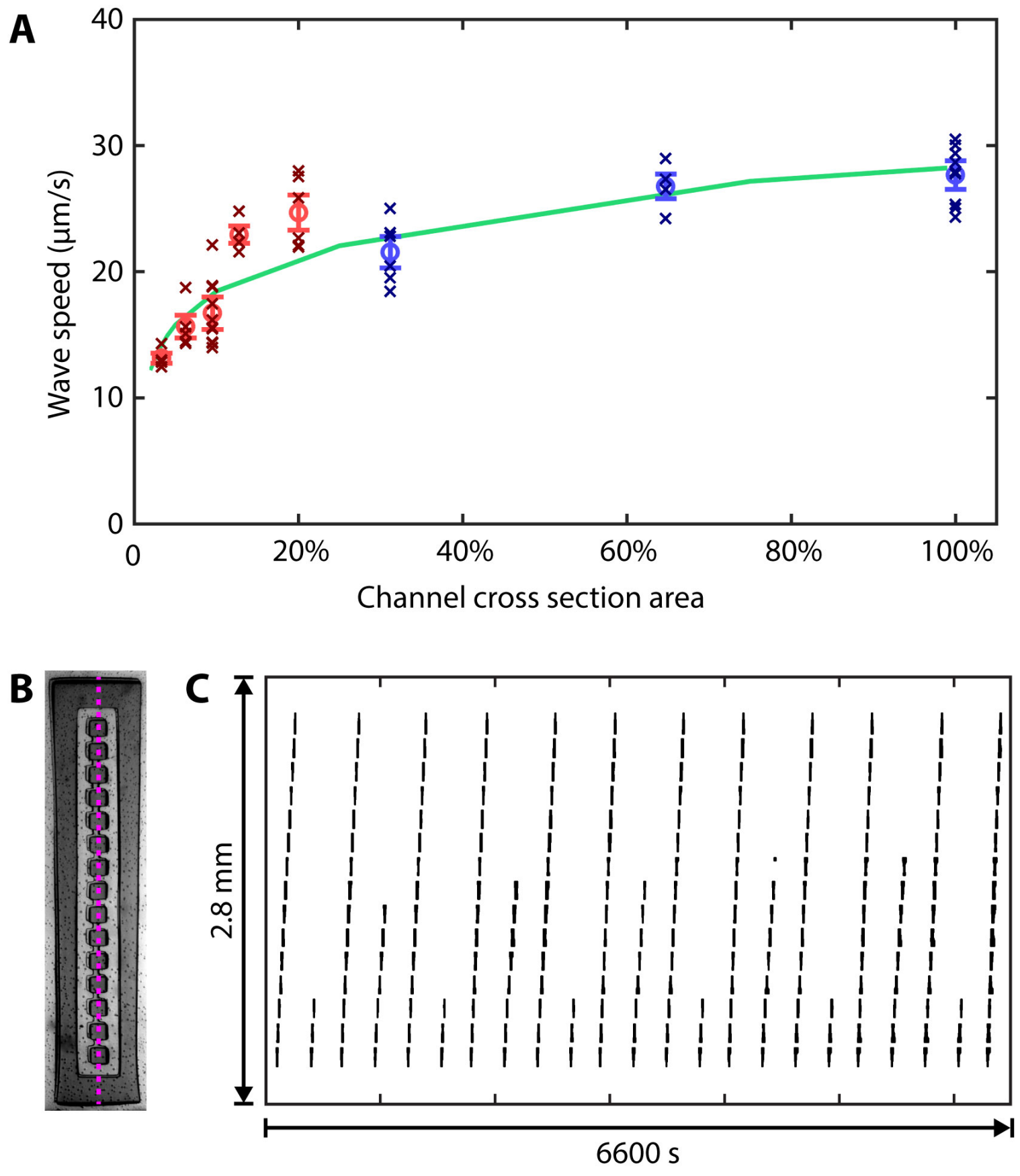

**Fig. 6. Channel size dependency of wave speed.** (**A**) Wave speed vs. channel cross section; 48 individual experiments. Well dimensions were constant; 100 µm x 100 µm x 110 µm (L x W x H). Red: channel width of 20 µm and varying channel height. Blue: channel width of 100 µm and varying channel height. Crosses: measurements, circles: average velocities, error bars: standard deviation. Green line: 2-D finite element simulations of wave speed vs channel cross section for the 2-D equivalent of the experiment (absolute wave speed is scaled down by a factor of 0.4 to match experiment). Photograph (**B**) and space-time plot (**C**) of chemical waves in a linear array with small channels of 6% cross-section. The pink line in (B) indicates the spatial component of the space-time-plot. Half of the waves terminate before reaching the end of the array. The termination location repeats every 4 cycles. The experiment is shown in movie S3.

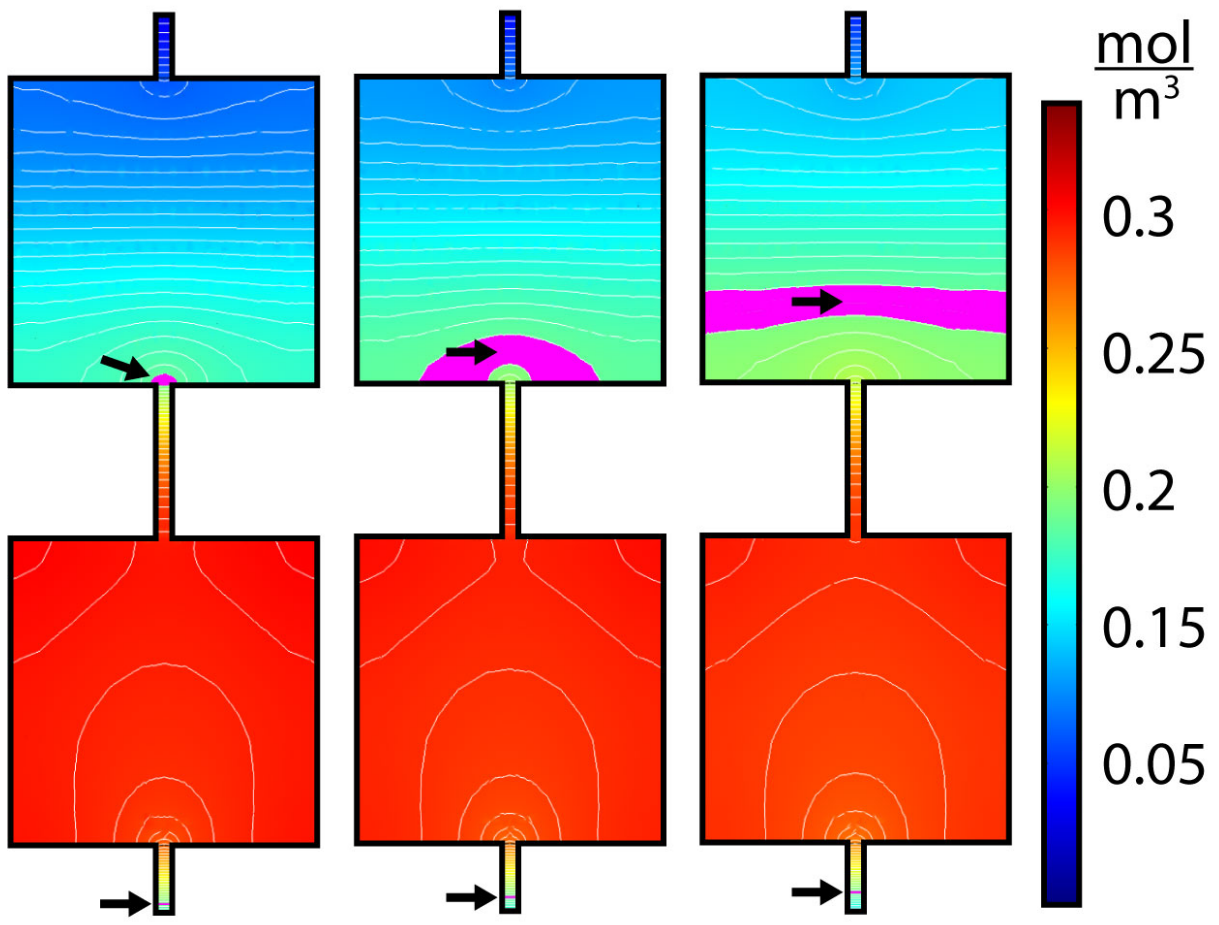

**Fig. 7. FEM simulation of wave propagating through connected wells.** Activator concentration (red: high, blue: low) calculated by FEM simulation at three, equally spaced times. Black arrows point to the magenta contour, arbitrarily chosen to be half the maximum activator concentration, which highlights the advancing (wide channel) and receding (narrow channel) wave fronts. The wave front in the wide channel advances much further than the wave front in the narrow channel over the same time intervals. See movie S4.

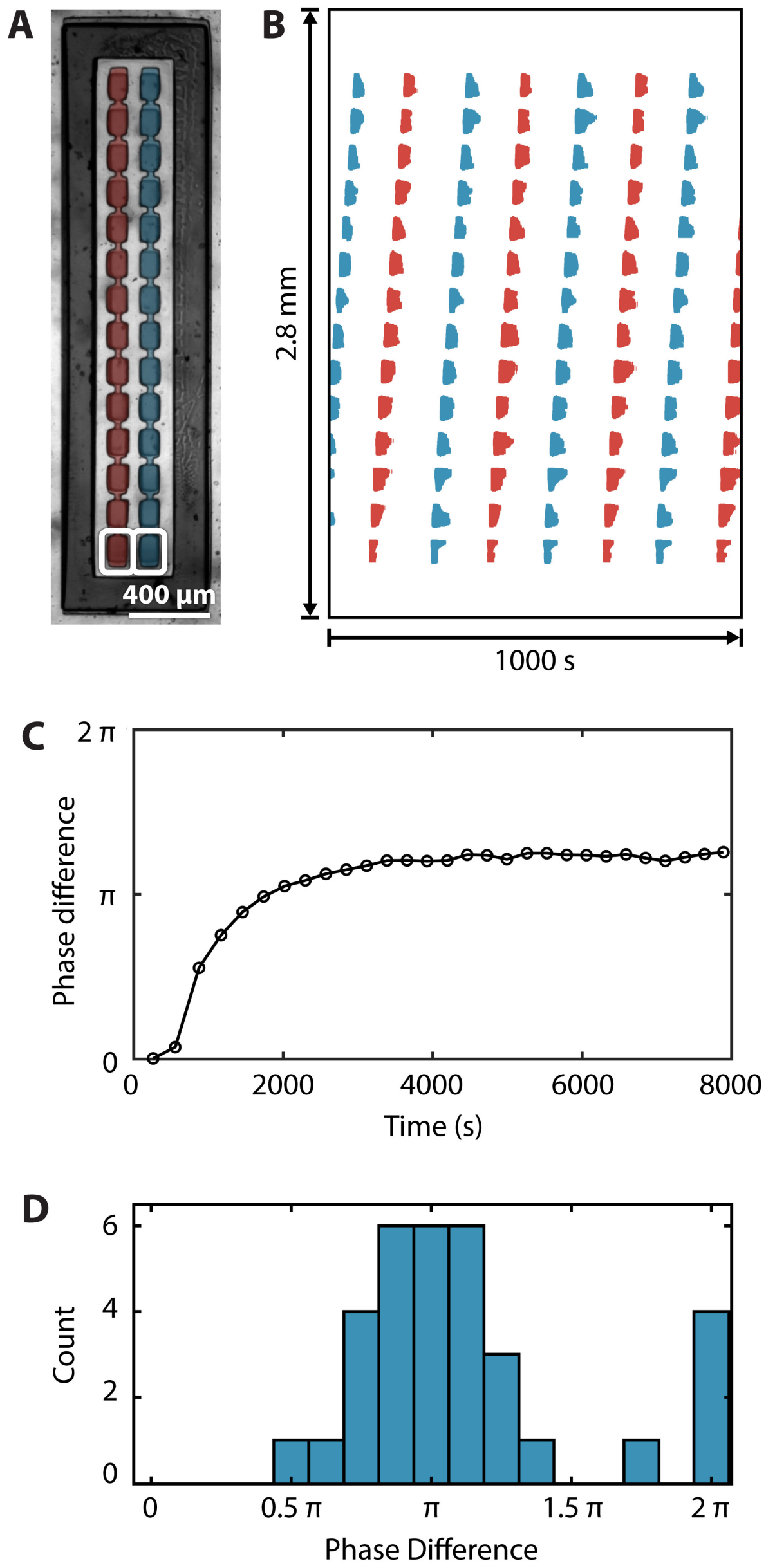

**Fig. 8. Bioinspired central pattern generator.**[2] (**A**) Two excitatory coupled linear arrays constructed from PDMS are fabricated side-by-side and filled with BZ solution. The arrays are surrounded by a BZ filled moat to maintain constant chemical boundary conditions. All wells are 120 μm length x 80 μm width x 90 μm height. Connecting channels are 50 μm long, 20 μm wide and 30 μm deep. The two rows of wells are separated by 70 μm of PDMS. Pacemaker wells,

uninhibited by light, are indicated by white boxes. See movie S6. (**B**) Space-time plots of the network shown in (A). Two space-time plots are superimposed. Red traces show oscillations of wells that are marked red in (A) (left column), blue traces show oscillations of wells marked in blue (right column). The rows of the space-time plot in (B) are aligned with the rows of the BZ array in (A). Traveling waves within a particular linear array of wells appear as parallel slanted red or blue lines in the space-time plot, indicative of a constant velocity. Oscillations in the two neighboring arrays are anti-phase. (**C**) Phase difference between pacemaker wells in (A) and (B) vs. time. The two arrays start with pacemakers in-phase and gradually evolve to an anti-phase pattern. Linear array separation: 70 μm; (**D**) Steady-state phase difference distribution for 33 experiments.

# Supplemental Materials for

## Engineering reaction-diffusion networks with properties of neural tissue

Thomas Litschel, Michael M. Norton, Vardges Tserunyan and Seth Fraden

**The PDF file includes:**

## Materials and Methods

### *Fabrication of photoresist master*

Photoresist masters for casting the final PDMS device are created with standard photolithography methods. Briefly, we spin coat SU-8 negative photoresist (MicroChem SU-8 2075) on a silicon wafer, soft-bake the photoresist prior to the UV light exposure, place and align a photomask with the desired features on the photoresist, UV-expose the photoresist, hard bake the photoresist, and develop the master with propylene glycol methyl ether acetate. We design the photomasks using Autodesk AutoCAD® and the masks are printed commercially (CAD/Art Services, Inc.). In devices with channels, we use two-height features to make interconnecting channels with lower heights than the wells. This requires two layers of photoresist, which were aligned using alignment-markers and partial development.[1] The schematic in Fig. S7A shows photomask, the silicon-photoresist master and the final PDMS chip (see next section).

### *Fabrication of PDMS microfluidics*

A schematic of the fabrication setup is shown in Fig. S1A and a video of the fabrication process is shown in movie S6. We mix 10 - 15 g of PDMS (Dow Corning SYLGARD® 184 Silicone Elastomer Kit) with a centrifugal mixer (Thinky® planetary non-vacuum centrifugal mixer). Before applying the PDMS to the silicon-photoresist master, we coat the master with Cytop® CTX-109AE at a dilution 1:40 for 5 seconds. Cytop® is an amorphous fluoropolymer which is used as a mold release agent and is essential for separating the PDMS and photoresist. We place the silicon-photoresist master in a desiccator and pour a small amount of PDMS (just enough to cover the microfluidic features) onto the master. We de-gas the PDMS in the desiccator for 10 minutes until the PDMS is bubble-free. Two 25 mm x 75 mm glass slides are plasma cleaned (plasma cleaner: Diener Zepto) and put onto the PDMS. We then place the setup onto a flat surface and cover it with a sheet of Mylar® to facilitate disassembly after curing. We place two more glass slides orthogonally to the carrier glasses on the stack. A lead weight of 15 kg is placed onto these glass slides. We let the PDMS cure for 12 hours at room temperature before replacing the heavy weight with a lighter weight and put the entire setup in a 70°C oven for another 6 hours. After curing, the setup is carefully disassembled. The master can be reused without re-applying Cytop®. The final microfluidic device is shown in Fig. S7B.

***BZ composition***

We use a photosensitive version of the Belousov-Zhabotinsky reaction that employs malonic acid as the organic substrate, ferroin redox indicator as a metal catalyst, and $[Ru(bpy)_3]^{2+}$ as a photosensitive co-catalyst. Other components of the reaction are sulfuric acid, sodium bromide and sodium bromate. The final BZ solution contains the following chemicals in the given concentrations: Ferroin (3 mM), sodium bromide (25 mM), malonic acid (400 mM), sulfuric acid (80 mM), sodium bromate (288 mM) and $[Ru(bpy)_3]Cl_2$ (1.2 mM). The components are mixed in the same order as they appear here. Stock concentrations can be found in table S1.

***Experimental protocol***

To maintain BZ oscillations for several hours and to achieve the desired coupling, the microfluidic wells must be properly sealed. The BZ reaction generates $CO_2$ gas that will create bubbles if open to air. Therefore, we seal the BZ solution air tight with no air bubbles in the wells so that the $CO_2$ gas remains dissolved in solution. Additionally, we seal the device to prevent the contents of the array of wells to come into contact with each other and with the BZ in the moat used to establish boundary conditions. To achieve these requirements, we designed a clamping device, which isolates the desired volume from air by a glass lid and applies pressure in a way that all wells are evenly sealed. The device consists of a bottom and top clamp frame that has been laser cut from Plexiglas plates. A rubber O-ring encloses an additional reservoir of BZ solution surrounding the sealed area to diminish concentration gradients and diffusion from the inside through the PDMS. The PDMS wells are sealed by a 3.2 mm x 3.2 mm x 1.7 mm glass window attached to a 25 mm x 25 mm x 1.2 mm glass slide. A round 10 mm in diameter and 3 mm in height PDMS window attached to another 25 mm x 25 mm glass slide allows for even pressure applied directly over the sample. This is important as otherwise the glass bends and lifts off the surface of the wells by a few microns, which is enough space to generate leaks. The entire setup is shown in Fig. S1B.

Before the BZ solution is loaded onto the PDMS chip, the chip is plasma treated in order to functionalize the PDMS surface and render it hydrophilic. Skipping this step prevents BZ solution from properly filling the wells due to the small size of the wells, the hydrophobicity of PDMS and the surface tension of the aqueous solution. We mix the 6 reagents of the BZ reaction together in aqueous solution and immediately pipette the solution onto an array of wells. We then place the setup into our clamp, and the device is tightened under a microscope so that a slight degree of compression is visible, which facilitates proper sealing. Before we start video recording, we wait for 30 to 40 minutes. This has shown to be crucial when choosing high initial concentrations of sodium bromide, since premature sample illumination can prevent oscillations from occurring. The process is shown in movie S7.

***Programmed illumination and recording***

For our experiments, we use a homemade programmable illumination microscope.[2] Its main component is a commercial three-color liquid-crystal display (LCD) projector with the optics inverted so that it reduces the image instead of expanding it as intended by the manufacturer. The computer projector is controlled by MATLAB code that is able to project light onto individual wells to control BZ oscillations by light inhibition.[2] To illuminate the sample for recording, we use a uniform Köhler illumination with a cyan LED and a green filter that filters light of a wavelength of 510 ±10 nm. Green light marginally excites $[Ru(bpy)_3]^{2+}$, but is well suited to distinguish the two states of the BZ reaction.[3] When imaged with a black-and-white CCD camera, the BZ solution appears bright in its oxidized state (blue) and dark in the reduced state (red).

## Experimental Details

***Experiments to quantify wave speed in dependency of channel size (Fig. 4, Fig. 5, Fig. 6)***

The experiments shown in Fig. 4, 5 and 7 and movies S1 and S3 use a one-dimensional array of wells with a moat-like structure surrounding the array. The "moat" provides defined boundary conditions, as illustrated in Fig. 3A. At the same time, we use the moat to calibrate the light intensity. As shown in Fig. 3B, we use light to create pacemaker wells. We shine the same intensity of light on all BZ compartments, including the moat, except for a single well at the end of the 1D array of wells. Within a small range of light intensities, the BZ solution in the surrounding moat will not oscillate, but we can observe waves traveling through the entire array of wells, originating at the pacemaker well.

It is important to note that if we omit the light inhibition, in many cases unidirectional waves still emerge. Fig. S4B shows the results of an experiment in which only the moat is light-inhibited, but not the one-dimensional arrays. Similar to the light controlled case (Fig. S4A), there is a clear dependence of wave speed on geometry. The propagation speed in the continuous compartment is about double the speed compared to that of the other arrays. Note that the propagation speed values in Fig. S4B are about three times as high as in the light controlled equivalents (Fig. S4A). However, these experiments without pacemaker wells are not reliably reproducible. Fig. S4 (C and D) show period times for both types of experiments.

While the wave speed is relatively constant over time, Fig. S4 (E and F) shows it is not constant across space within traveling waves. Close to the boundaries of the array (the beginning and the end of the array), we observe an increased propagation speed in experiment and simulation.

All experiments use an array of fifteen 100 μm x 100 μm x 110 μm (L x W x H) wells connected by channels that are 50 μm long. The channels connecting the wells in the experiment shown in Fig. 4, Fig. 5 and movie S1 are 20 μm wide and 52 μm deep. The continuous chamber is 100 μm wide, 2200 μm long and 110 μm deep. In the experiment shown in Fig. 6 (B and C) and movie S3 channels are 20 µm wide and 34 µm deep. In all experiments the moat is 150 µm wide and separated from the linear arrays by 80 µm of PDMS.

***Bioinspired central pattern generators (Fig. 8)***

The experimental procedure is similar to the procedure for the experiments in Fig. 8. Instead of having a single pacemaker well, here we have two uninhibited pacemaker wells, located next to each other at one end of each column with excitatory links. To assure in-phase starting conditions, we initially shine light on all wells for the duration of about one BZ oscillation period, which is of order 300 seconds. All wells are 120 µm x 80 µm x 90 µm (L x W x H). Connecting channels are 50 µm long, 20 µm wide and 40 µm deep. The moat is 150 µm wide. 70 µm of PDMS separates moat and the arrays.

## Simulations

We model the BZ reaction using the Field-Körös-Noyes (FKN) model.[4] The model tracks the evolution of six key chemical intermediates of the BZ cycle (bromous acid $x$, bromide $y$, oxidized catalyst $z$, bromine dioxide radical $w$, hypobromous acid $p$) and assumes that the feed stocks of the reaction (bromate $a$, malonic acid $m$, proton concentration $h$, and bromomalonic acid $b$) remain constant.

$$
\begin{aligned}
\frac{dx}{dt} &= -k_1 xy + k_2 y - 2k_3 x^2 - k_4 x + k_r w^2 + k_{red} wc \equiv R_x(\mathbf{C}) \\
\frac{dy}{dt} &= -k_1 xy - k_2 y - k_5 yp + k_6 u + k_7 u + k_9 z + k(I)\frac{cb}{b_c + b} \equiv R_y(\mathbf{C}) \\
\frac{dz}{dt} &= k_{red} wc - k_9 z - k_{10} z + k(I)\frac{cb}{b_c + b} \equiv R_z(\mathbf{C}) \\
\frac{dw}{dt} &= 2k_4 x - 2k_r w^2 - k_{red} wc \equiv R_w(\mathbf{C}) \\
\frac{dp}{dt} &= 2k_1 xy + k_2 y + k_3 x^2 - k_5 yp + k_6 u - k_8 p \equiv R_p(\mathbf{C}) \\
\frac{du}{dt} &= k_5 yp - k_6 u - k_7 u \equiv R_u(\mathbf{C})
\end{aligned}
\qquad [1.1]
$$

The values for the rate constants and concentrations we use are listed in table S2. Additional source terms $\propto k(I)$ added to the rate equations for the inhibitor and catalyst include the effect of the light sensitive catalyst $[Ru(bpy)_3]^{2+}$. While our solution contains both Ferroin and $[Ru(bpy)_3]^{2+}$ catalysts, we do not make a distinction between them in our model of the system.

### *Finite Element Model (FEM)*

To model the reaction-diffusion front propagation through the linear arrays, we use the finite element software COMSOL to solve the full reaction diffusion equations in two-dimensions

$$\frac{\partial \mathbf{C}}{\partial t} = \mathbf{R}(\mathbf{C}) + D\nabla^2 \mathbf{C} \ , \qquad [1.2]$$

where $\mathbf{C}=\{x\cdots u\}$ and the reactions $\mathbf{R}(\mathbf{C})$ are as stated in Eq. 1.1. At the interface between the aqueous phase and the PDMS a Robin's boundary condition for the Bromine in each phase maintains the partitioning between the phases as follows

$$\begin{aligned} \hat{\mathbf{n}}\cdot D_{u,aq}\nabla u_{aq} &= k\left(Pu_{aq}-u_P\right)\Big|_{\partial\Omega_{aq/P}} \\ \hat{\mathbf{n}}\cdot D_{u,P}\nabla u_P &= -k\left(Pu_{aq}-u_P\right)\Big|_{\partial\Omega_{aq/P}} \end{aligned} \quad . \qquad [1.3]$$

The parameter $P\sim 2.5$ is the PDMS - aqueous partition coefficient for bromine and $k$ is a large constant that enforces the partition coefficient while ensuring mass flux continuity across the interface; past simulations[5] have used values $O(10^4)$, we find that computational times can be reduced by using values $O(10^2)$ without materially altering the solution. The principal assumption here is that the interface is always locally at equilibrium and therefore does not offer any resistance to mass transfer. At the boundaries of the computational domain and at the PDMS - aqueous interface for all other species, the natural no-flux boundary condition $\hat{\mathbf{n}}\cdot D_C\nabla\mathbf{C}=0\Big|_{\partial\Omega}$ is used. We used two levels of light inhibition a high value $O(10^{-3})$ in the moat to maintain the constant chemical boundary condition and a lower level $O(10^{-6}\text{-}10^{-5})$ to push the chemistry from the oscillatory regime into the excitable state.

## Supplementary Figures

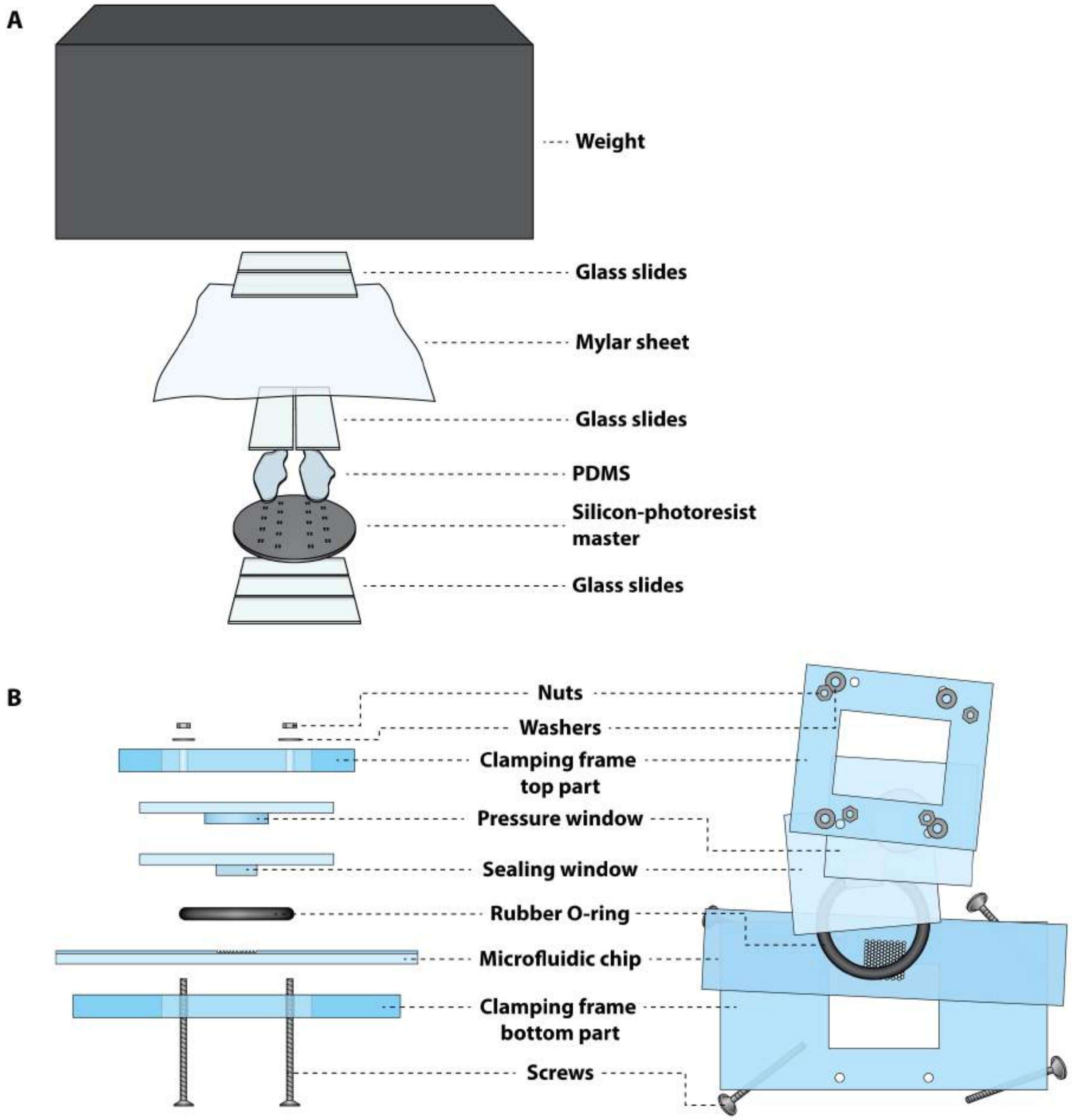

**Fig. S1.** (**A**) Schematic of setup for fabricating a thin PDMS microfluidic device. From bottom to top: (1) Three glass slides to provide an even surface. (2) Silicon-photoresist master. (3) PDMS that is poured onto the features of the master. (4) Plasma cleaned glass slides as base of the microfluidic chip. (5) Sheet of Mylar to facilitate disassembly. (6) Glass slides to distribute pressure of weight. (7) 15 kg lead weight. Fabrication process is shown in movie S6. (**B**) Experimental Setup. Side view (left) and top view (right) of disassembled clamping device used for sealing a PDMS chip loaded with BZ solution. From bottom to top: Screws, bottom acrylic glass clamping frame, patterned microfluidic chip, rubber O-ring, sealing

glass window, pressure glass window, top acrylic glass clamping frame, washers, screw nuts. Objects are not to scale. For simplification purposes the depicted microfluidic chip contains only one array of wells; the actual microfluidic chips contain 15 sections that can be individually sealed (see Fig. S7). The assembly of the clamping device shown movie S7.

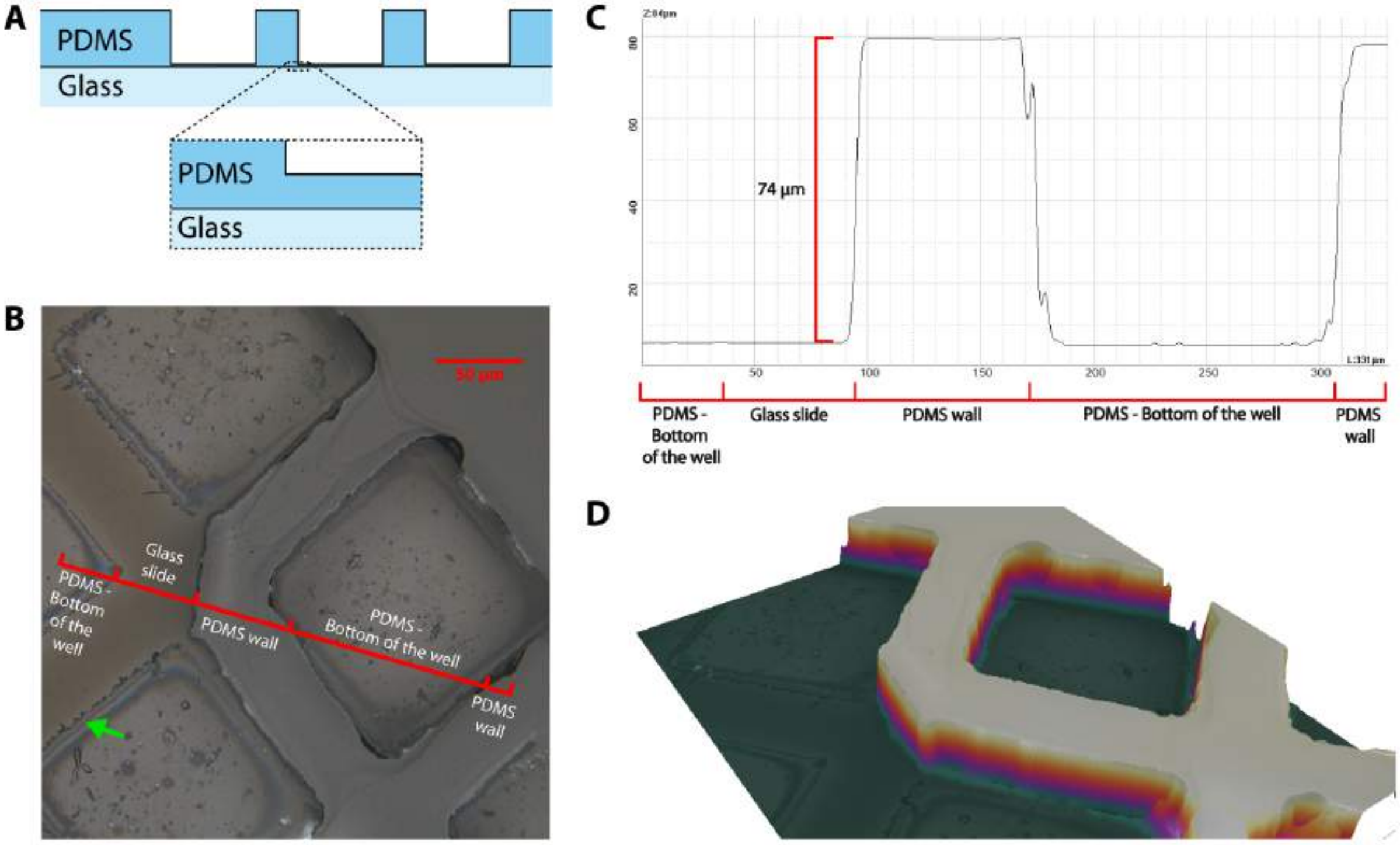

**Fig. S2.** (**A**) Section view of microfluidic device. Magnified area illustrates thin layer of PDMS at the bottom of each microfluidic well. (**B**) Photograph of a microfluidic chip with parts of the PDMS sheet removed with a scalpel: PDMS walls on the left side are missing. The thin layer of PDMS at the bottom of each well remains attached to the glass. Interference fringes (green arrow) are visible which indicate that the thickness of the PDMS on the bottom of the well is under 1 micron thick. (**C** and **D**): Topographical rendering of microfluidic features, with partially removed PDMS layer. (C) 2D profile of the section corresponding to the red line in the top view image shown in (B). PDMS layer at the bottom of wells is too thin to be visible in the profile. (D) Rendered 3D image of the topological scan. Color indicates height.

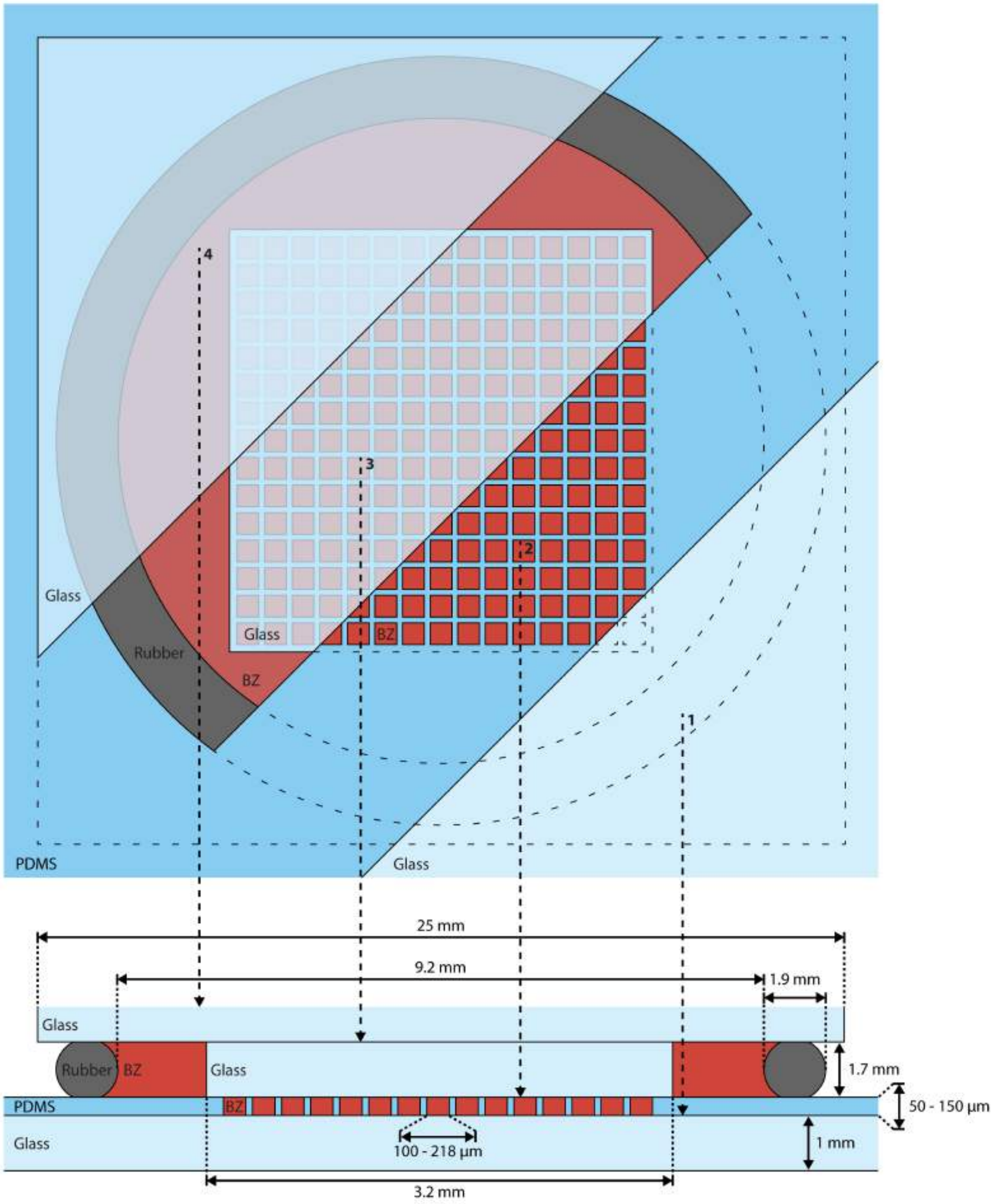

**Fig. S3.** Schematics of BZ-loaded microfluidic features. Top: Tiered schematic of microfluidic setup. From bottom right to top left the following layers are laid bare: (1) Glass slide as the base of the microfluidic chip. (2) PDMS sheet in which wells are patterned. Wells are filled with the BZ reaction solution. (3) Glass window that seals the microfluidic wells, surrounding BZ solution and rubber O-ring that confines the BZ solution. (4) Glass slide to which sealing glass window in *3* is attached. Dimensions are not to scale. Bottom: Section view of microfluidic setup with actual dimensions labeled.

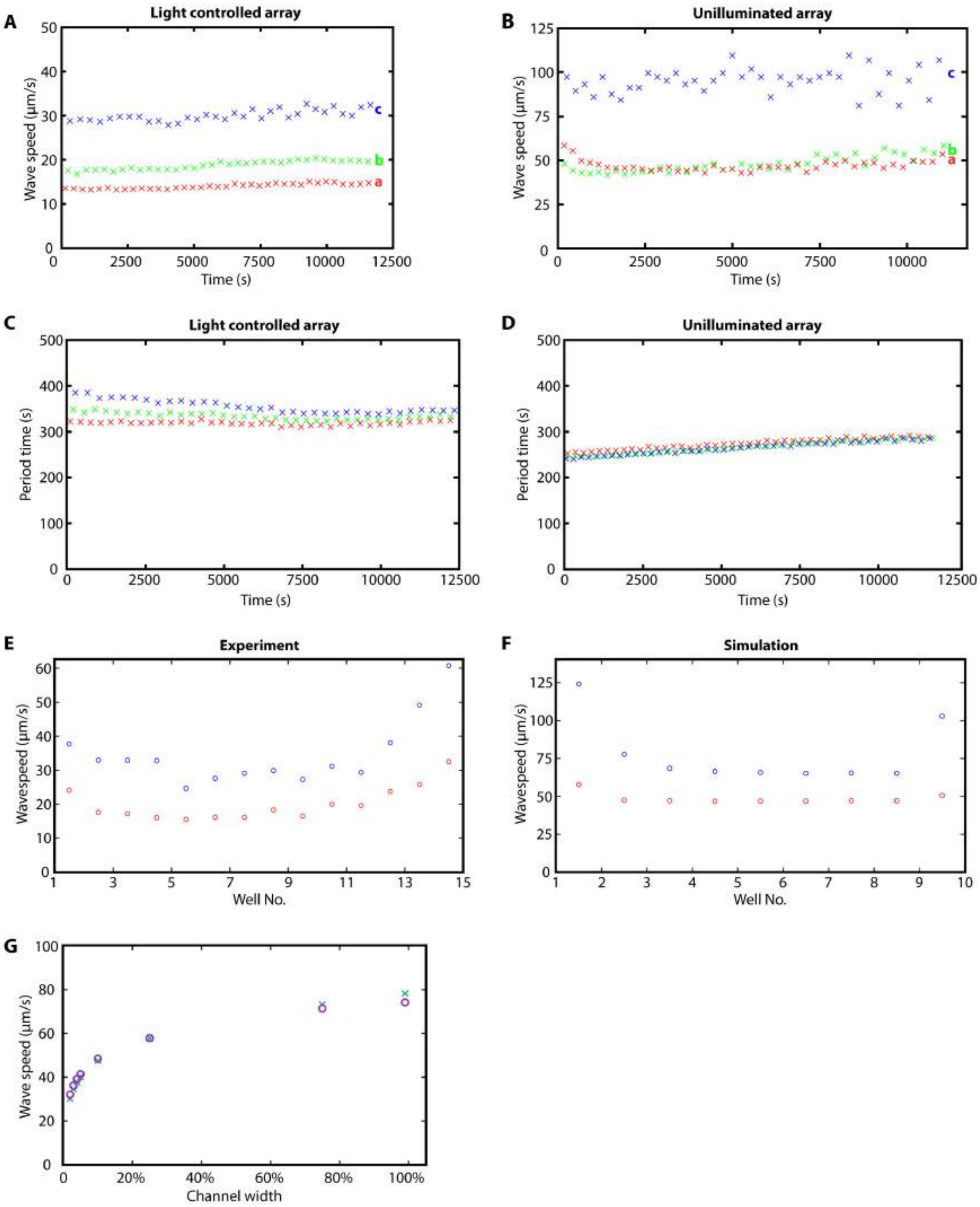

**Fig. S4.** (**A**) Propagation speed of waves in the microfluidic design shown in Fig. 4C as a function of time. Each color represents the wave speeds in one of the three columns (Fig. 4C). (a) and (b) are the two geometrically identical arrays, consisting of fifteen wells connected by channels. (c) is a continuous chamber, which can be considered as an array with wells and channels of equal size. Each data point is the average speed of a single propagating wave. The greatest wave speed is in the continuous channel (c). The difference in wave speeds between (a) and (b) shows typical experimental variances for identical geometries. (**B**) Propagation speed of waves in experiment without controlled pacemaker. As in (A), each data point is the average speed of a single propagating wave and the letters (a), (b) and (c) represent the same columns from the PDMS device shown in Fig. 4C. In contrast to the wave speed plot in (A), here we didn't shine light on the arrays of wells, but only inhibited the BZ filled moat with light. The greatest wave speed is still in the continuous chamber (c), but wave speed is scaled up. (**C** and **D**): Period times of oscillations. (C) Periods of oscillations of the pacemaker well of each of the three one-dimensional arrays as seen in Fig. 4C over the course of the experiment. Colors of the data points match the colors in (A) and (B). (D) Same plot for the experiment without pacemaker wells (B). Shown are the period times in the wells in which the waves originate. (**E** and **F**): Spatial dependency of propagation speed. (E) Spatial dependency of the propagation speed in experiment with array of wells (design like Fig. 4C). As the chemical wave travels from the first well in the array to the last well, the propagation speed of the wave is not constant. In red is the wave speed of an array of 15 wells with a channel cross section area of 9.5% of the entire cross section of a well. In blue the wave speed of a continuous chamber representing an array of wells with 100% channel cross section area. (F) Plotted are the same properties, but from a 2D simulation of the same kind of experiment with 10 wells in a row. In red is the propagation speed of wells that are connected by channels that have 10% of the width of a well; in blue the propagation speed of wells that are connected by channels that have 99% of the width of a well. We found that in experiments and in simulations the wave speed at the beginning and the end of an array is higher compared to regions in between the two. (**G**) 2-D finite element simulations of wave speed vs channel cross section for BZ waves through linear array of wells. Addition to Fig. 6A: Fig. 6A shows simulation accounting for diffusion through PDMS as green line, which is shown here as purple circles. Simulation with no flux conditions are shown as green crosses and show very little difference in wave speed.

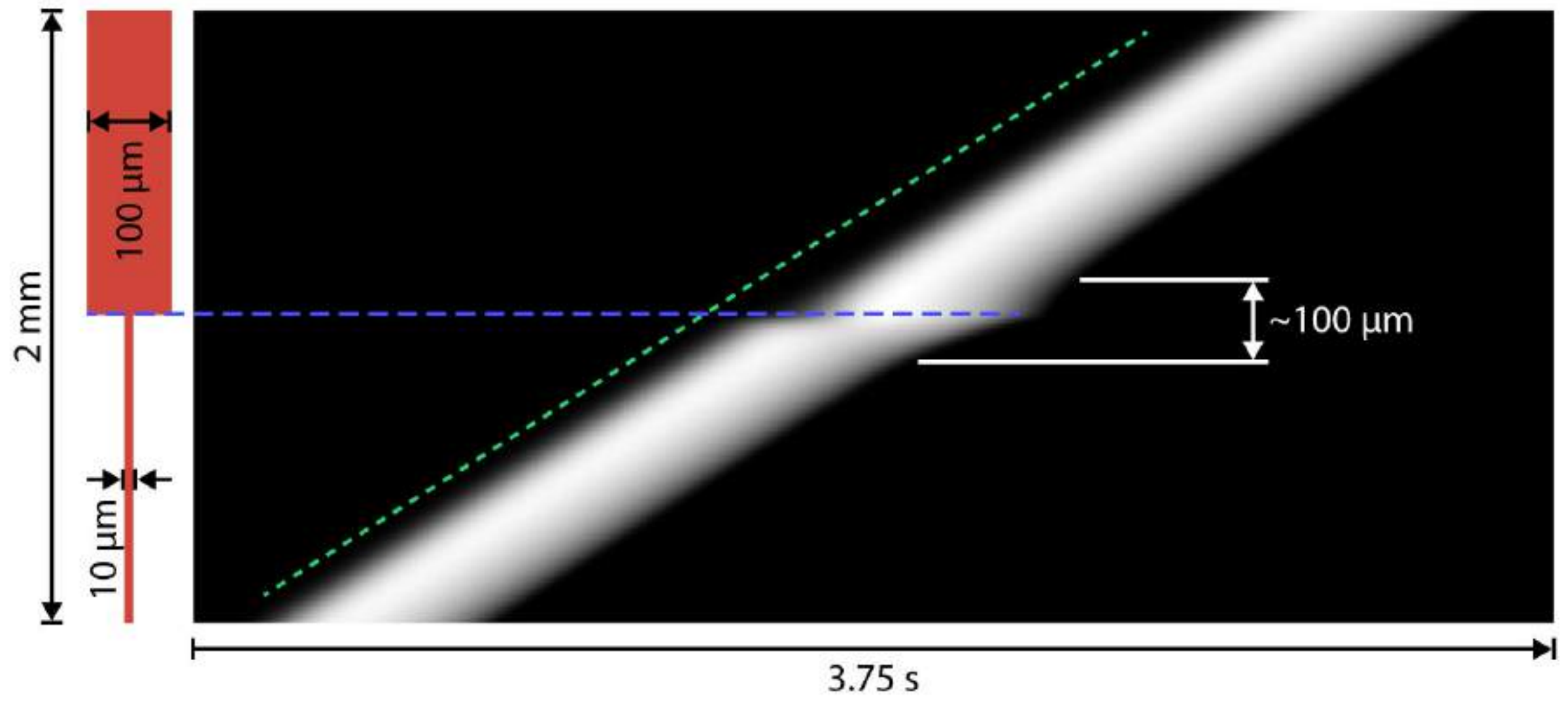

**Fig. S5.** Space-time plot (right) of FEM predictions showing the propagation of the activator front in a channel with no-flux boundary condition that undergoes a step-change in width (left). The slope of the plot (dashed green line) corresponds to the wave speed. The plot shows that the wave speeds far from the step-change in channel widths are identical, indicating that only the change in width, not the magnitude of the width, determines wave speed. We are also able to garner that the upstream and downstream spatial influence of the step is on the order of the wide channel's dimension, 100 µm.

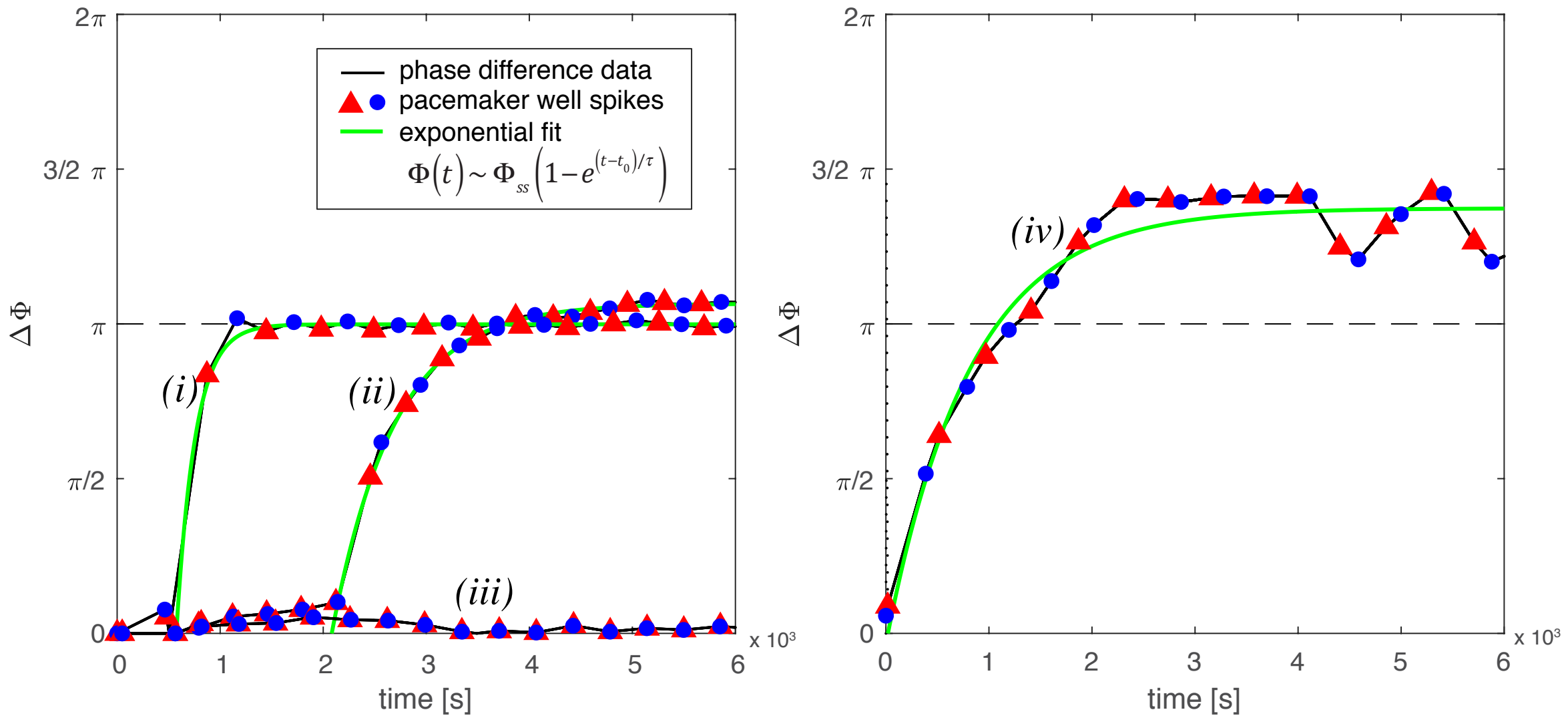

**Fig. S6.** Phase difference dynamics for four different experiments exhibiting different behaviors: (*i*) rapid convergence to steady state after a few nearly in-phase oscillations, (*ii*) slightly slower convergence to steady after multiple in-phase oscillations, (*iii*) in-phase for the duration of the experiment, and (*iv*) slow exponential approach immediately after light is removed and convergence to an undesired steady state. In (*i*) and (*ii*) only the data after the initial in-phase oscillations have ceased is used for the exponential fit. The dynamics of cases (*i*) and (*ii*) exhibit dynamics captured by our finite element analysis (Fig. S8), which also predicts that the lag time depends on coupling strength. Fully characterizing the coupling time experimentally is a task we leave for future work; however, we summarize our current findings in Fig. S7.

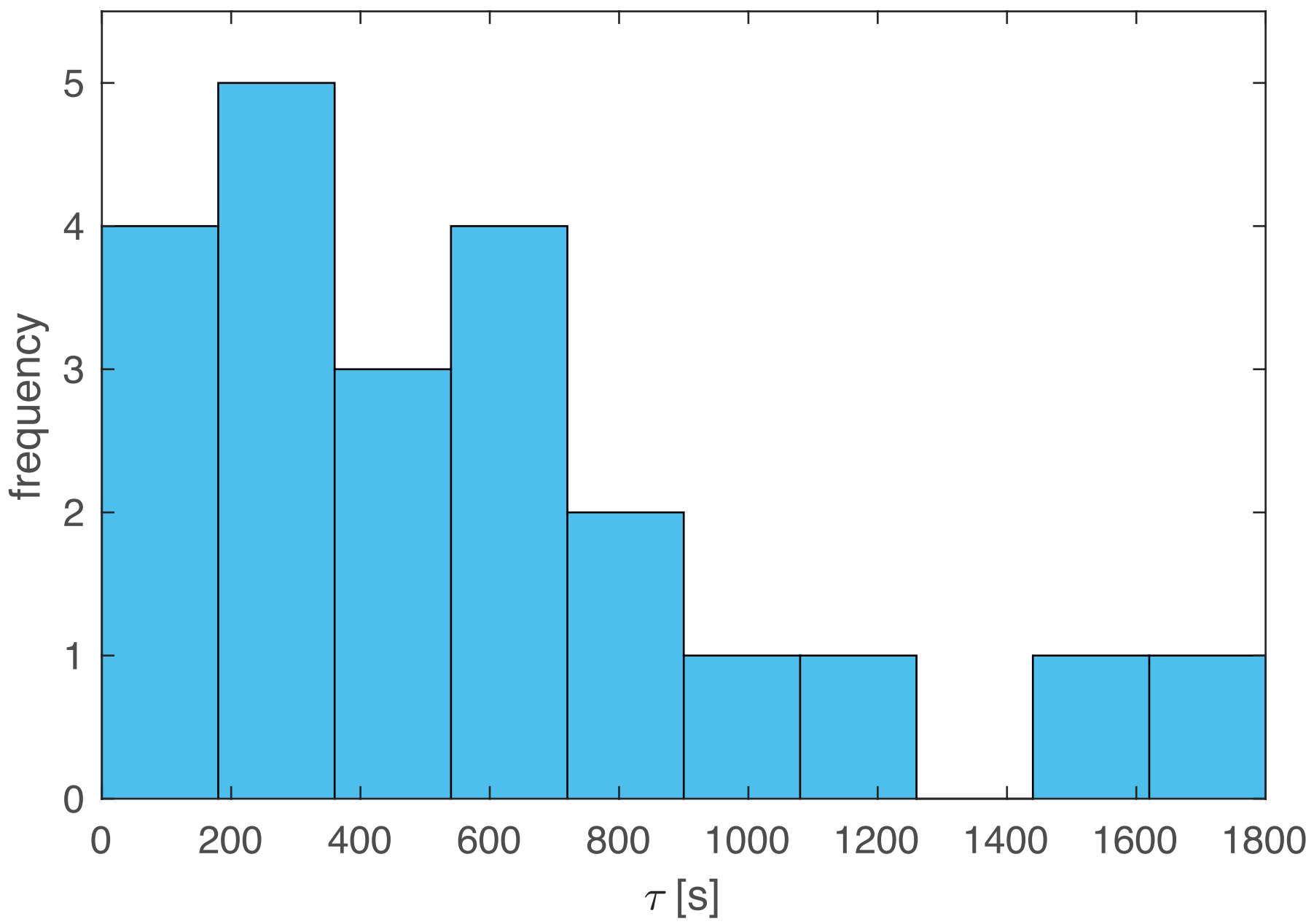

**Fig. S7.** To characterize the strength of the dynamical attractor, we plot a histogram of coupling time $\tau$ measured by fitting an exponential of the form $\Delta\Phi(t) \sim \Delta\Phi_{ss}(1 - e^{-(t-t_0)/\tau})$ to experimental data. Distance between excitatory channels varies from 50 to 70 microns. In-phase oscillations (see Fig. S6) are excluded from the fit. The distribution is peaked around 300 seconds but is broad. For comparison, we explore coupling dynamics over a large range of distances in simulation (Fig. S8) but do not find a broad distribution in the exponential part of the coupling time. We suspect that chip-to-chip variations in the PDMS layer beneath the wells (nominally 1 micron) contributes to the broadness of this distribution; a theory that remains to be tested.

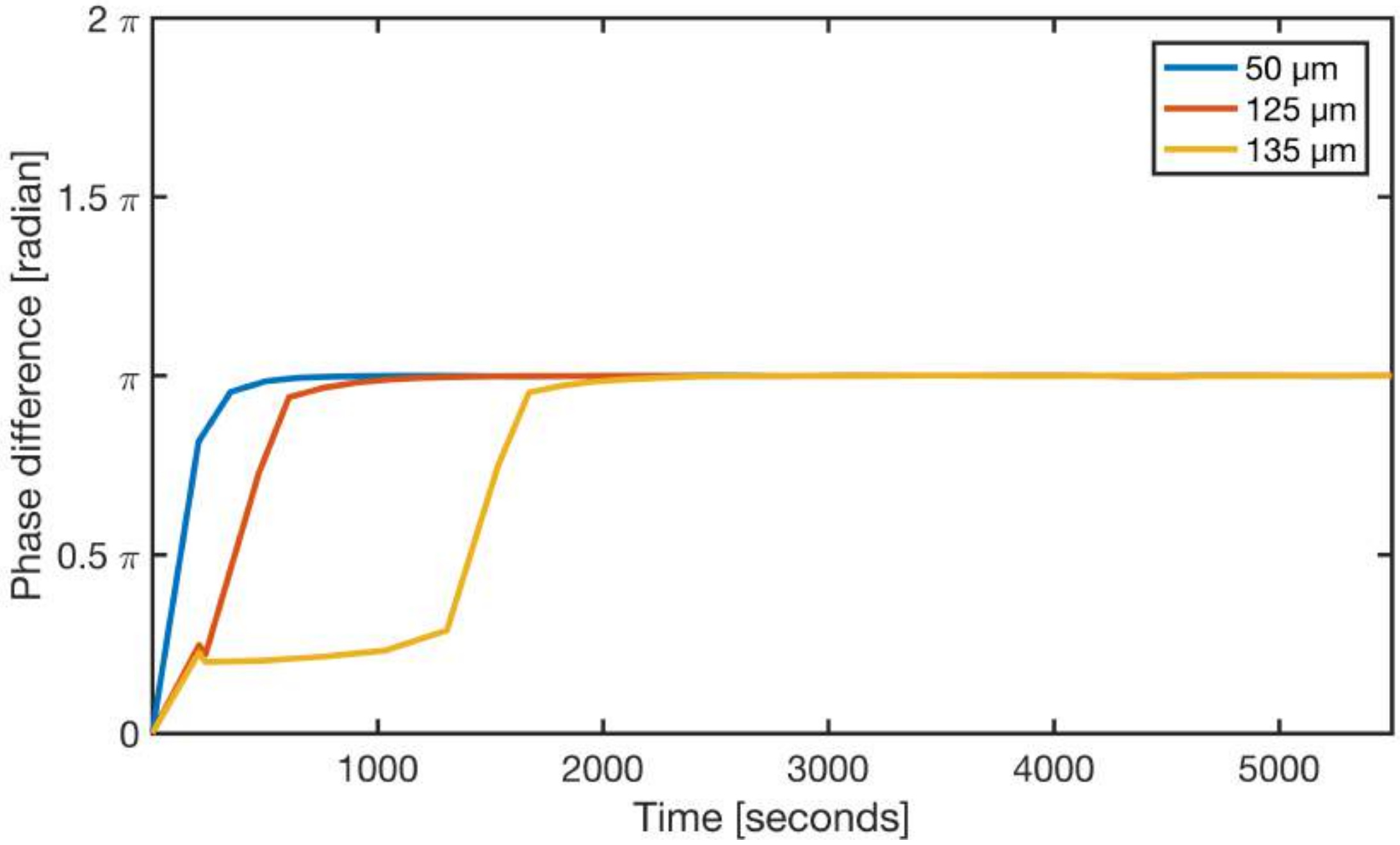

**Fig. S8.** FEM simulation of BZ-CPGs. Phase difference between the linear arrays vs. time. The two arrays start with pacemakers near in-phase and switch to an anti-phase pattern. Shown is the evolution of phase difference for three different separation distances: 50 μm, 125 μm, and 135 μm. Interestingly, varying the distance increased the lag time while minimally impacting the exponential part of the approach to steady state. Since we were unable to produce the long coupling times ($\tau$ > 800 s) observed in experiment, we posit that experimental factors such as the thickness of the PDMS layer beneath the wells impacts dynamics in certain regimes, a hypothesis that would require a full 3D simulation to explore theoretically.

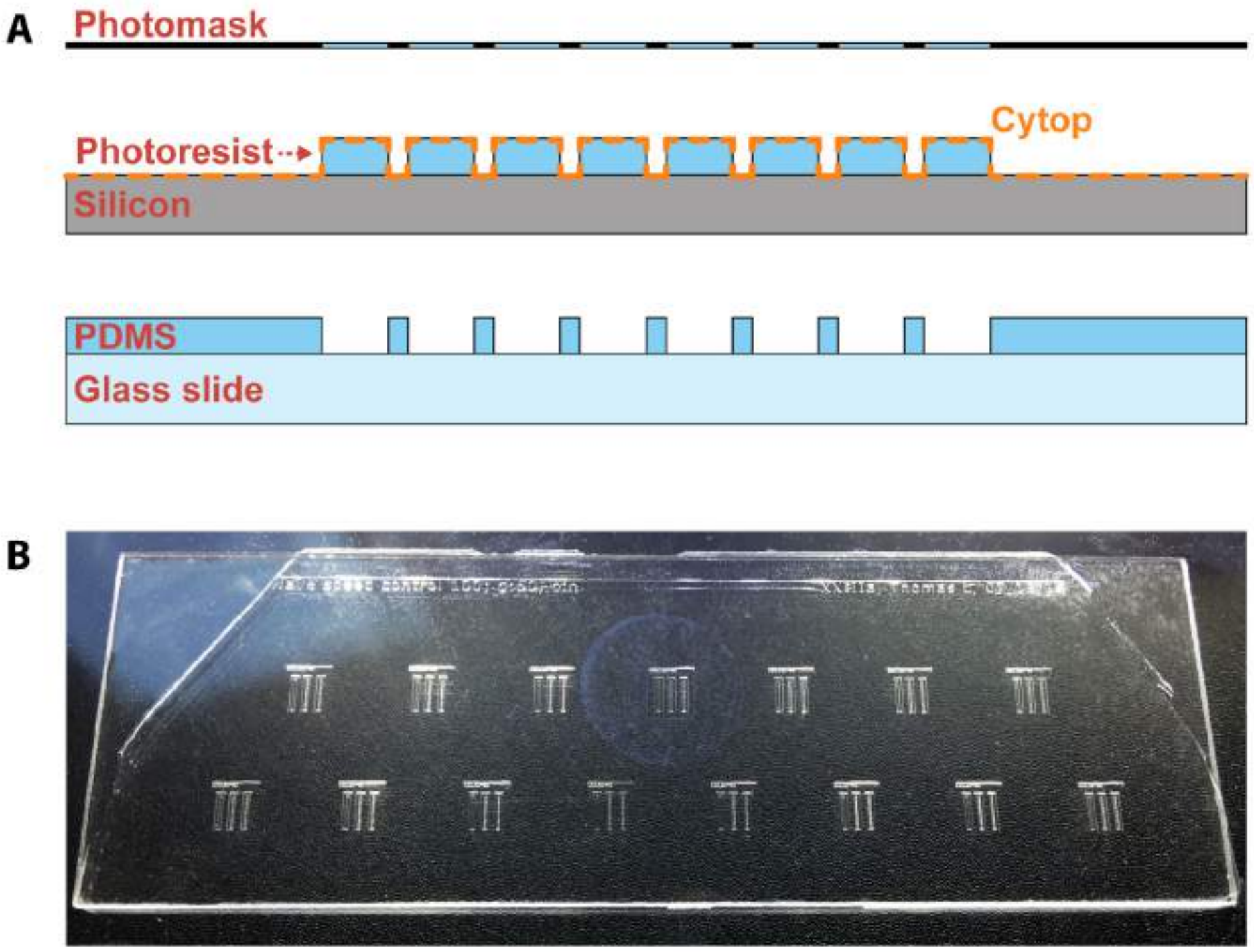

**Fig. S9.** (**A**) Illustration of the stages of the fabrication process. From top to bottom: Photomask, silicon-photoresist master and final chip. Structures are not to scale. The fabrication of the silicon-photoresist master and the final microfluidic device are described in the methods section. (**B**) Photograph of a microfluidic chip. Each chip contains 15 independent devices of which each one can be used in an experiment after being sealed by a 3.2 mm x 3.2 mm glass window. On the depicted chip, each device contains three 1D arrays similar to the ones in Fig. 4C.

## Supplementary Tables

| **Name** | **Chemical Formula** | **Final Concentration** | **Stock Concentration** |
|---|---|---|---|
| Malonic Acid | $CH_2(COOH)_2$ | 400 mM | 2400 mM |
| Sulfuric Acid | $H_2SO_4$ | 80 mM | 480 mM |
| Sodium Bromide | NaBr | 25 mM | 125 mM |
| Sodium Bromate | $NaBrO_3$ | 288 mM | 1728 mM |
| Ferroin | $C_{36}H_{24}FeN_6^{2+}$ | 3 mM | 18 mM |
| $[Ru(bpy)_3]Cl_2$ | $C_{30}H_{24}N_6Cl_2Ru{\cdot}6H_2O$ | 1.2 mM | 7.2 mM |

**Table S1.** BZ reagents. List of BZ reagents with common names, chemical formulas, final concentrations and stock concentrations. To prepare the BZ reaction solution with the stated final concentrations, we mix equal volumes of stock solutions of each reagent with the specified stock concentrations.

| Species | Symbol | Value | Rate Constant | Value |
|---|---|---|---|---|
| $[HBrO_2]$ | x | ~ | $k_1$ | $2\times10^6 [M^{-2}s^{-1}] h$ |
| $[Br_-]$ | y | ~ | $k_2$ | $2 [M^{-3}s^{-1}] h^2 a$ |
| $[Br_2]$ | u | ~ | $k_3$ | $3000 [M^{-1}s^{-1}]$ |
| $[Fe(phen)_3^{3+}]$ & $[Ru(bibpy)_3^{3+}]$ | z | ~ | $k_4$ | $42 [M^{-2}s^{-1}] ha$ |
| $[Fe(phen)_3^{2+}]$ & $[Ru(bibpy)_3^{2+}]$ | c | ~ | $k_5$ | $5\times10^9 [M^{-2}s^{-1}] h$ |
| $[BrO_2^\bullet]$ | w | ~ | $k_6$ | $10 [s^{-1}]$ |
| $[HOBr]$ | p | ~ | $k_7$ | $29 [M^{-1}s^{-1}] m$ |
| $[H^+]$ | h | $160\ [mM]$ | $k_8$ | $9.3 [M^{-1}s^{-1}] m$ |
| $[BrO_3^-]$ | a | $288\ [mM]$ | $k_9$ | $0.12 [M^{-1}s^{-1}] m$ |
| $[CH_2(COOH)_2]$ | m | $400\ [mM]$ | $k_{10}$ | $0.05 [M^{-1}s^{-1}] m$ |
| $[BrCH(COOH)_2]$ | b | $0.12m$ | $k_r$ | $2\times10^8 [M^{-1}s^{-1}]$ |
| $[Fe(phen)]+[Ru(bibpy)]$ | $c_0$ | $4.2\ [mM]$ | $k_{red}$ | $5\times10^6 [M^{-1}s^{-1}]$ |

**Table S2.** BZ species in simulations. Symbol key, reaction rates and constants used in the FKN[4] (Eq. 1.1)

.

## Supplementary Movies

**Movie S1:** Waves traveling in three 1D reaction-diffusion geometries are shown. The left and middle arrays consist of fifteen wells connected by channels. The rightmost is a continuous chamber, representing an array with wells and channels of equal size. The rows of wells are surrounded by a large compartment, or "moat", to establish constant chemical boundary conditions. The moat is filled with BZ solution and illuminated with light to suppress oscillations. The left and middle arrays are identical and consist of fifteen 100 μm x 100 μm x 110 μm wells connected by channels that are 20 μm wide, 50 μm long and 52 μm deep. The long column on the right is a 100 μm wide, 2200 μm long and 110 μm deep chamber. On the left array, light is shone on all wells except for the well at the bottom. The one uninhibited well acts as a pacemaker. The pacemaker for the middle array is the top well and the pacemaker for the rightmost, continuous channel is located at the bottom. A space time plot of the left array is shown in Fig. 5B.

**Movie S2:** FEM simulations of wave propagation through a linear array. Wells are 100 μm x 100 μm, and connected with 80 μm long, 5 μm wide channels. 0:00-0:14: Simulation showing an entire array of 15 wells; color shows the concentration of oxidized catalyst. 0:14-0:28: Close-up with two wells connected by a channel. FEM simulation showing activator concentration (color map), flux (arrows) and the geometry of the wave front (a contour corresponding to half the maximum activator concentration was labeled with a thick, magenta line to help visualize the wave) as it emanates from a small channel.

**Movie S3:** Experiment like in movie S1, but with channels that are only 34 μm deep. Half of the waves terminate before reaching the end of the array. A space time plot of this video is shown in Fig. 6C.

**Movie S4**: Two excitatory coupled linear arrays constructed from PDMS are placed side-by-side and filled with BZ solution. The arrays are surrounded by a moat. All wells are 120 μm x 80 μm x 90 μm. Connecting channels are 50 μm long, 20 μm wide and 40 μm deep. 0:00-0:25: Experiment with a design in which 70 μm of PDMS separates the two linear arrays (shown in Fig. (A to C)). 0:25-2:08: Experiment in which only 50 μm separate the two linear arrays.

**Movie S5:** FEM simulation of dual channel CPG, color shows the inhibitor (bromine) concentration which drives the columns towards antiphase synchrony. The dimensions of each linear array are the same as those in movie S3; the columns are separated by 50 μm.

**Movie S6:** Fabrication of a thin PDMS device. See methods section and Fig. S1A.

**Movie S7:** Loading a PDMS device with the BZ solution and sealing procedure. See methods section and Fig. S1B.

## Supplementary References